\documentclass[twocolumn]{aastex62}

\usepackage{amsmath, amssymb}
\usepackage{graphicx}
\usepackage{xcolor}
\usepackage{hyperref}
\usepackage{enumitem}
\hypersetup{colorlinks=true,allcolors=blue}
\usepackage{comment}
\usepackage{placeins}       

\newcommand{\kms}{\,\mathrm{km\,s^{-1}}}

\shortauthors{Chouliaras et~al.}

\begin{document}

\title{Automatic detection of Flare Ribbon Fine Structures as Proxies for Plasmoid Dynamics in Flare Reconnection}

\correspondingauthor{Georgios Chouliaras}
\email{georgios.chouliaras@durham.ac.uk}

\author{Georgios Chouliaras}
\affiliation{Department of Mathematical Sciences, Durham University, Durham, DH1 3LE, UK\\
}

\author{Peter F. Wyper}
\affiliation{Department of Mathematical Sciences, Durham University, Durham, DH1 3LE, UK\\
}
\author{Joel T. Dahlin}
\affiliation{Heliophysics Science Division, NASA Goddard Space Flight Center, Greenbelt, MD, USA\\
}
\affiliation{Astronomy Department, University of Maryland, College Park, MD 20742, USA}

\author{Lyndsay Fletcher}
\affiliation{Scottish Universities Physics Alliance, School of Physics and Astronomy, University of
Glasgow, Glasgow, UK \\
}
\affiliation{Rosseland Centre for Solar Physics, University of Oslo, Oslo, Norway}

\begin{abstract}
Flare ribbons often display fine structures along their fronts that are commonly interpreted as signatures of intermittent reconnection dynamics  
including plasmoid formation in the flare current sheet.
We introduce an automated method that detects and tracks the spiral- and wave-like imprints of these structures and as a proof of concept apply it to maps of magnetic field-line length from a high-resolution 3D eruptive-flare simulation. The workflow applies the correlation-dimension method, density-based clustering, and a minimum-area ellipse fit to summarise each feature. We show that across the simulated flare, 
the detected spirals remain locked to the ribbon's outward motion while drifting coherently along the ribbon. The two ribbons show opposite along-ribbon drift and motion away from their hooks in accordance with theoretical expectations, with instantaneous speeds of $10$--$800$ km s$^{-1}$, all well below the local surface Alfv\'{e}n speed. Occurrence, lifetimes, and mean magnetic flux of the features peak during the impulsive phase. The distribution of per-spiral mean unsigned flux shows a scale-free tail above roughly $6\times10^{18}$ Mx with a power-law exponent near $3.4$. Together, these results show that bursty, plasmoid-mediated flare reconnection leaves a clear, measurable signature on the flare ribbons. 
The method provides a practical surface diagnostic of ribbon fine structure that can potentially be used to inform our understanding of three-dimensional magnetic reconnection in the flare current sheet. 
\end{abstract}


\section{Introduction}\label{sec:intro}
Solar flare ribbons are the chromospheric and transition region signature of magnetic reconnection in solar flares. They are thought to arise from the deposition of energy into the chromosphere via energetic particles, thermal conduction, or even Alfvén waves associated with the magnetic reconnection process at the heart of the flare \citep{FletcherHudson2008ApJ,Battaglia2009AA,Holman2011SSR}. As such, flare ribbons and the flare reconnection process are intrinsically linked. 

At large scales, the overall morphology of flare ribbons and its links to the reconnection process are relatively well understood. Ribbons appear in pairs

at the conjugate foot points of newly formed magnetic field lines. The surface magnetic flux swept out by the ribbons is therefore often used to infer the rate of reconnection within the flare \citep{Qiu2004ApJ_ReconnectionRate,Miklenic2007_AA,Kazachenko2017ApJ_RibbonDB}. Furthermore, depending on the complexity of the magnetic field in and around the flaring region these ribbon pairs can have a variety of morphologies. 

By far the most commonly studied is the case where a flare occurs in conjunction with an erupting flux rope/coronal mass ejection. In this context, the ribbons form roughly 
parallel to the polarity inversion line along which the erupting structure originated. These parallel sections were the focus of the original ``standard" 2D CSHKP flare model \citep{Carmichael_1964,Sturrock_1966,Hirayama_1974,Kopp_etal1976}. More recent work has extended the process to 3D where, amongst other things, the often observed ``hooks" or ``J-shapes" at the end of the parallel sections in the ribbons of these flares are now understood to be wrapping around where the erupting flux rope is anchored to the surface \citep[see e.g.][ for recent reviews]{Kazachenko2022,Dudik2025}. More complex topologies involving pre-existing coronal null points, and more generally quasi-separatrix layers are associated with more complex ribbon morphologies, including quasi-circular ribbons with associated remote ribbon counterparts \citep[e.g.][]{Zhang2024}. And indeed, erupting flux ropes embedded within such topologies have multiple associated ribbons, with the location and evolution of the different ribbons 
directly linked to the reconnection occurring between the erupting magnetic structure and its surroundings in the corona \citep[e.g.][]{Wyper2021,Wyper2024,Kumar2021,Karpen2024}. 

Alongside this progressively better understanding of the global context and 3D nature of ribbon formation there has also been a significant improvement of our understanding of 3D reconnection in general. It is now well accepted theoretically that high aspect ratio current sheets (with lengths much greater than their width in the inflow direction) 
of the type expected in flares are highly unstable to the tearing/plasmoid instability at Lunquist numbers typical of the solar corona \citep[e.g.][]{Shibata_etal2001,Loureiro2007,Wyper2014a,Huang2016,Beg2022}. This implies that the flare reconnection process is intrinsically bursty, occurring within a fragmented, potentially turbulent reconnection region. Such regions are expected to be dominated by the presence of plasmoids that take the form of small-scale flux ropes, typically associated with denser plasma, that evolve within and are then ejected from the reconnection region \citep{Guidoni2011DensityEnhancements,Liu2013PlasmoidEjections,Takasao2016PlasmoidMotions}. This picture of bursty flare reconnection is consistent with for example post-CME blobs \citep{Song2012_SolPhys_CMEBlobs,Patel2020_AA_PostCMEPlasmoids}, super-arcade downflows \citep{McKenzieHudson1999_ApJL_SADs,Savage2012_ApJL_SADsReinterpretation} and spectral broadening \citep{Milligan2011_ApJ_NonthermalBroadening,Warren2018_ApJ_CurrentSheetSpectroscopy}, suggesting flare reconnection is indeed intrinsically bursty in nature. It also suggests that there should be an imprint of this bursty reconnection on the flare ribbons themselves in the form of sub-structure within the ribbon, and further that such sub-structure could potentially provide a diagnostic of the flare reconnection process. 

Observationally, there are a growing number of examples of ribbon sub-structure.  
For instance, recently \citet{Faber2025} identified and analysed a chain of regularly spaced blobs 140\,km to 200\,km in size in Swedish Solar Telescope data which they attribute to fragmentation of the flare current layer. Similarly, \citet{Jing2016} find similar scale ranges of 80\,km to 200\,km for blob-like structures within ribbons in Goode Solar telescope data. The latest results from the Daniel K. Inouye Solar Telescope \citep[DKIST;][]{rimmele20a} also show fine-scale structure in both the newly formed coronal loops and the ribbons themselves \citep[e.g.][]{Tamburri2025,Yadav2025}. \citet{Lorinckik2025} recently found rapidly moving bright features with speeds greater than $1000$\,km/s. On larger scales, the X-class flare SOL2014-09-10T17:45 was captured with AIA and IRIS has been the focus of study of several authors as it showed rapidly evolving, spiral and wave-like features  \cite[e.g.][]{LiZhang2015ApJL,Dudik2016ApJ,Wyper_Pontin2021ApJ...920..102W,CorchadoAlbelo2024}.

One prominent interpretation for, in particular, fast moving blobs is that they represent the heated foot points of field lines which are rapidly slipping through, and being energised in the flare current sheet as a result of the 3D nature of the reconnection process. The motion of the blobs is apparent, and occurs as field lines sequentially slip in the coronal region, with the speed related to the gradient in the field line mapping \citep{Janvier2013A&A...555A..77J}. 
Such slippage can explain fast moving blobs along the ribbon, and is often supported by sequential brightening of coronal loops in EUV connected to such blobs \citep[e.g.][]{Dudik2016ApJ}, but it struggles to explain wave-like and spiral ribbon features. 
 
In their analytical model, \citet{Wyper_Pontin2021ApJ...920..102W} showed that the presence of plasmoids in the flare current layer should naturally imprint wave-like and spiral shapes upon flare ribbons in addition to the slipping reconnection associated with the overall flare reconnection. The magnetic twist within the plasmoids naturally wraps the up the flux surfaces that connect to the ribbons, imprinting the spirals. As with slipping reconnection, the wrapping motion is an apparent one generated by the change of magnetic connectivity within and around the plasmoids. Key predictions from this model are that (1) in eruptive flares the chirality of spirals/perturbations matches that of the hooks, 
(2) plasmoids map to between two and four sub-structures within the ribbons depending upon their location and orientation within the current layer, and (3) that the spiral/wave-like features should drift along the ribbon away from the hooks in the straight sections, and towards the end of the hooks in the hooks themselves. Recently, \citet{Dahlin2025} presented a high-resolution adaptive mesh magnetohydrodynamic simulation of an eruptive, two-ribbon flare that confirmed the qualitative predictions of the simplified analytical model, but further showed the dynamics and interaction of the plasmoids and spirals over time. Another key finding from this study was that the wrapping of multiple plasmoids into one another creates significantly more internal structure within spirals \citep[something also noted in the context of 3D null point reconnection in][]{Wyper2014a}.
In the work of \citet{Wyper_Pontin2021ApJ...920..102W} and \citet{Dahlin2025} we laid out the general idea and demonstrated that the spiral and wave-like features suggested by the analytical model do form dynamically in fully dynamic simulations, while also outlining the qualitative ribbon and plasmoid dynamics. In this work we further analyse the simulation results 
of \citet{Dahlin2025}, focusing on a more quantitative and statistical analysis of the ribbon fine structures from the simulation. 

In the simulation of \citet{Dahlin2025} considered here, the plasmoids form initially near the center of flare current layer once it reaches a sufficiently high-aspect ratio, consistent with the tearing/plasmoid instability. However, we emphasize that flares can involve additional instabilities and structured dynamics (including shear-driven vortical motions such as Kelvin--Helmholtz-type behaviour, e.g. \citet{Karpen_etal2012,Che2020}). Our aim in this paper is not to argue for exclusivity of any one process, but to introduce an automated, quantitative method that detects and tracks the \emph{geometrically complex} portions of a projected ribbon-front (or analogous connectivity-proxy boundary). Because the method is based on boundary geometry rather than a specific physical trigger, it is broadly applicable to complex ribbon-front boundaries formed under a range of physical conditions.

We first introduce a new analysis code developed for the task, before presenting some initial results which we contrast against results from 3D plasmoid mediated reconnection theory and the latest observations. In Section 2 we summarise the simulation setup and the data we use for our analysis. In Sections \ref{sec:data} and \ref{sec:method} we outline the method for identifying and tracking spiral features and its application to this dataset. We then present our results in Section \ref{sec:results} and discuss our findings and future directions in Sections \ref{sec:discussion} and \ref{sec:conclusions}.

\section{Model and data description}
\label{sec:data}
\subsection{MHD Model}
The data utilized in this study were obtained from high-resolution, three-dimensional adaptive mesh refinement (AMR) ideal magnetohydrodynamic (MHD) simulation described comprehensively in \citet{Joel_2022aApJ...932...94D}. The initial magnetic field configuration consists of a multipolar active region embedded in a global dipolar background field, forming a breakout topology. Magnetic shear is introduced gradually at the lower boundary using the STITCH method \citet{Stitch2022ApJ...941...79D}, wherein tangential flux is injected to simulate the helicity buildup commonly observed in active regions prior to eruptive flares. This procedure leads to the formation of a filament channel characterized by highly sheared field lines oriented along the polarity inversion line (PIL).

Once a critical amount of shear flux is accumulated, the system undergoes an eruptive instability triggered by breakout reconnection, leading to the formation and eruption of a flux rope. The simulation accurately captures the formation of a vertical current sheet beneath the erupting flux rope, and achieves a high effective Lundquist number such that the current sheet is highly fragmented with the presence of multiple plasmoids. More importantly, the simulation also captures the essential features of two ribbon flares whereby newly reconnected flux in the simulation sweeps out two parallel ribbons upon which fine-structure associated with the plasmoids is imprinted. This makes this simulation an ideal test bed for developing and automating the identification of ribbon fine-structures.

\begin{figure*}[p]
\centering
\includegraphics[width=\textwidth,height=.42\textheight,keepaspectratio]{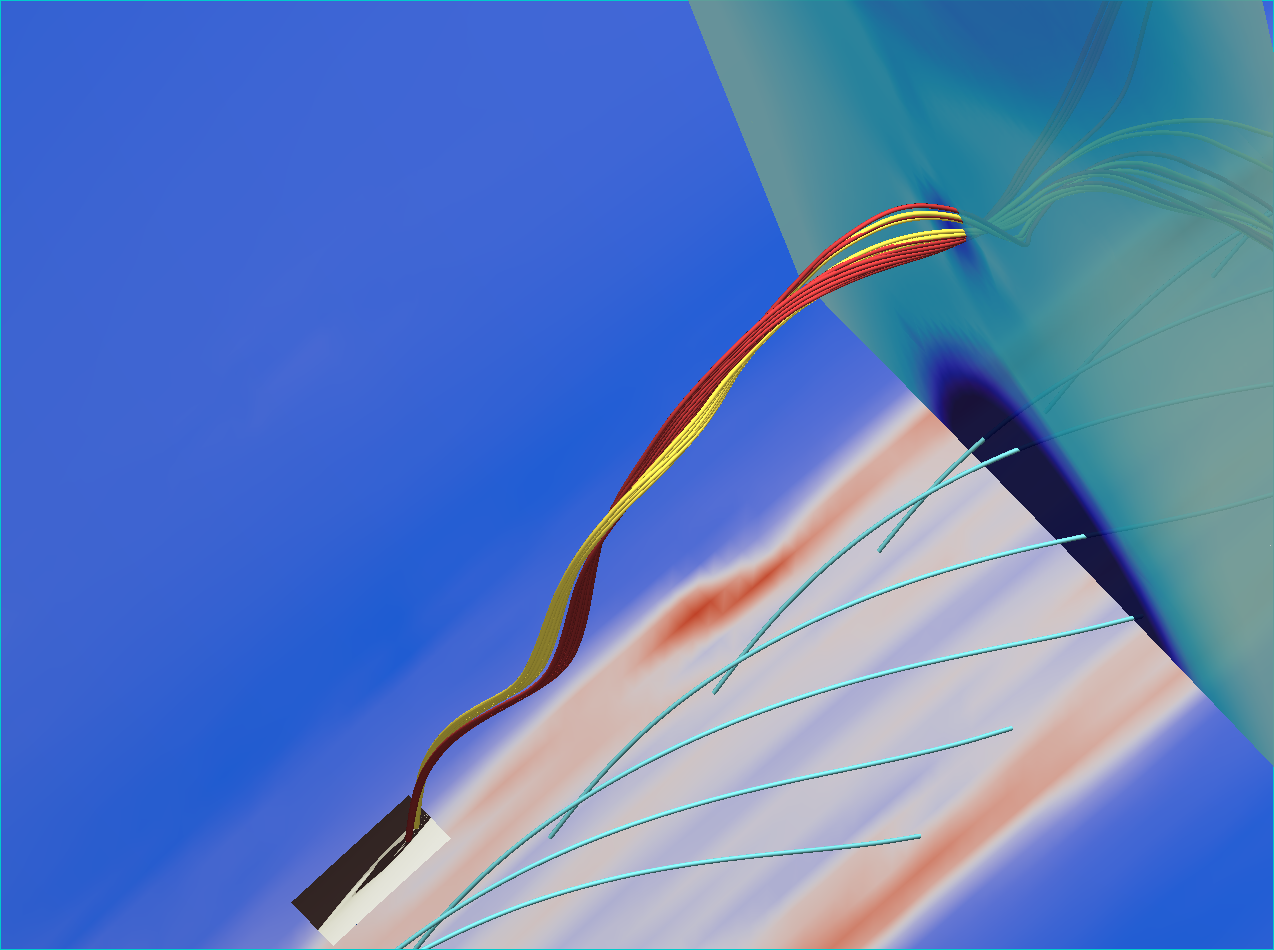}\vspace{0.6ex}
\includegraphics[width=\textwidth,height=.42\textheight,keepaspectratio]{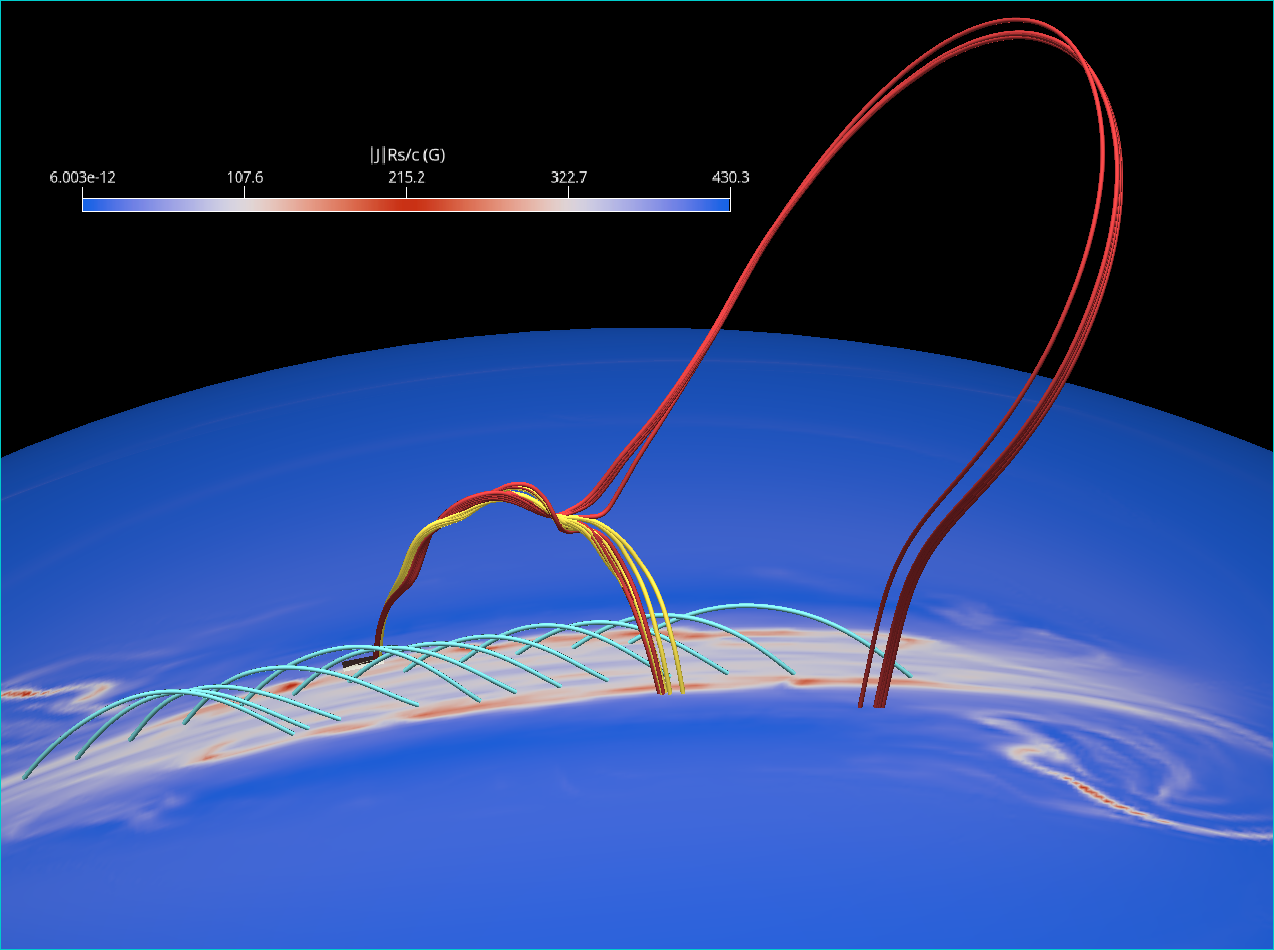}
\caption{Two complementary views of the same magnetic structure from the MHD run.
\textbf{Top:} Oblique close–up of the current sheet (semi–transparent teal slice, coloured by the guide field $B_{\phi}$).  
The black and grey insert shows the field line length for field lines traced from the lower boundary, indicating the ribbon front (the boundary between black and grey) has a local spiral feature. A bundle of field lines (red and yellow) show that field lines within the spiral map to a plasmoid within the flare current sheet.  
For reference, cyan lines trace the surrounding arcade rooted along the polarity–inversion line (PIL).
Bottom: Wider context 
view of the same bundle, showing that the plasmoid–threaded lines are also part of longer field lines of the erupting field.  
On the surface we display the current density distribution. }
\label{fig:plasmoid_spiral}
\end{figure*}

\subsection{Field-Line Length Maps}

As discussed in \citet{Dahlin2025} field line length maps (L-maps) present a simple and effective method for identifying the ribbon fronts in this simulation. To generate L-maps we utilize a fourth-order Runge-Kutta field-line tracing algorithm developed for the ARMS code \citep{wyper2016ApJ...820...77W,wyper2016ApJ...827....4W}. Starting from a dense grid of seed points on the photospheric plane, the magnetic field lines are numerically integrated until they reach the predefined domain boundaries or converge to their opposite photospheric endpoints. The total length of each traced field line is recorded and mapped back onto the initial seed point location. This systematic process results in two-dimensional maps that effectively represent the spatial distribution of magnetic field line lengths. L-maps are a fast proxy for magnetic connectivity. Each pixel’s value is  the arc-length of the field-line that threads it, so neighbouring pixels that share the nearby coronal foot-points display similar L. Connectivity jumps correspond in L-maps to abrupt changes in field line length.
Those sharp gradients outline flare ribbons, flux-rope hooks, spine lines and, crucially, the fine-scale spirals produced by plasmoids.

In the simulation of \citet{Joel_2022aApJ...932...94D} and in the analytic model of \citet{Wyper_Pontin2021ApJ...920..102W} every plasmoid that forms inside the flare current sheet is, topologically, locally, a small flux rope. Typically, these flux ropes forming in the flare current layer straddle QSL boundaries between pre and post-reconnection field lines. As a result, some parts of the plasmoid map to footpoints much farther along the ribbon than others, so the field–line length \(L\) is large there, whereas neighbouring parts can map nearby and have small \(L\). The hyperbolic flux tube (HFT) associated with the global QSLs defining the erupting flux rope and flare loops effectively splits this mapping, sending adjacent field lines to widely separated locations. The resulting ridge of large \(L\) values winds along the ribbon front, forming a compact spiral-like wrapping up of the boundary between short and longer field lines, hence, plasmoids imprint spiral-like features on the ribbon fronts shown in \(L\)-maps.
This is why they serve as the foundational diagnostic for our automated detection method. 

In our data, we demonstrate this connection in Fig.~\ref{fig:plasmoid_spiral}. The top panel shows a perpendicular slice through the current sheet, coloured by the guide magnetic field $B_{\phi}$. Plasmoids within the sheet produce compact enhancements in $B_{\phi}$ along this slice. The cyan bundle of field lines is traced from footpoints around the polarity–inversion line (PIL), which belongs to the flare loop. The red and yellow field lines are traced from a cross-section of a plasmoid, their helical geometry reveals magnetic twist. The plasmoid itself is highlighted by where the red and yellow field lines wrap around each other well away from the lower boundary. When these plasmoid field lines are mapped to the lower boundary (``surface''), their footpoints
organise into a spiral-like pattern. The inset (grey/black) shows the corresponding $L$-map on the surface, which exhibits a spiral co-located with those footpoints. The black colour corresponds to the longer field lines (that loop over the erupting flux rope) and the grey colour indicates the shorter ones newly formed by the flare reconnection.  
Furthermore, the bottom panel highlights the plasmoid’s complex connectivity: the field lines threading it are part of the erupting flux system, and the  HFT produces a clear splitting of the footpoint mapping. 

Typically, the imprint of the plasmoids on the ribbon front is of the form discussed above. But in general these features may appear as partial spirals, wave-like distortions or more complex but still localised features in the ribbon front \citep{Wyper_Pontin2021ApJ...920..102W}. Therefore, throughout this paper we will use the term ``spiral'' in an operational sense. We define a spiral as being a compact portion of the extracted ribbon-front 
whose geometry is locally folded, curled, or wound around a local centre, rather than being a locally smooth one-dimensional edge. As such, spiral features need not form a complete analytic spiral, nor do they need to follow any particular analytic form such as an Archimedean, logarithmic, or exponential spiral. The term is therefore morphological and operational, rather than a statement about the mathematical structure of the feature. However, this terminology is physically motivated as such features arise from field lines within the plasmoids spiralling around each other as they map to the surface.

\section{Overview of the automated method}
\label{sec:method}
\subsection{Spiral detection}
Our approach is split into two main phases. Automatic detection of ribbon sub-structure, and sub-structure characterisation. The first phase involves two main steps, the application of the Correlation Dimension Method  \citep[CDM,][]{Mason2022ApJ...937L..19M} and then the application of the Density-Based Spatial Clustering of Applications with Noise \citep[DBSCAN, ][]{DBSCANEsterKriegelSanderXu1996} to the result. Here we provide an overview of these steps in a general sense before discussing how the method was fine-tuned for this particular dataset.
The CDM is an established technique from nonlinear dynamics, introduced by \citet{GrassbergerProcaccia1983PhysicaD} and generalised by \citet{Mason2022ApJ...937L..19M} for use on coronal hole boundaries. It has also recently been applied to an observed flare ribbon in the context of better understanding variations in the flare reconnection rate \citep{CorchadoAlbelo2026}. Here we use it to evaluate the fractal dimensionality of boundary contours extracted from the processed L-maps. Mathematically, CDM involves calculating the correlation integral, $C(r)$, given by:
\begin{equation}
    C(r) = \frac{1}{N(N-1)}\sum_{i=1}^{N}\sum_{j \neq i}^{N} \Theta(r - \|\mathbf{x}_i - \mathbf{x}_j\|),
\end{equation}
where $\Theta$ is the Heaviside step function, $r$ is the scale of interest, and $\{\mathbf{x}_i\}_{i=1}^{N}$ are coordinates of points along boundary contours. For structures exhibiting self-similarity (fractality), $C(r)$ scales according to a power-law relationship:
\begin{equation}
    C(r) \propto r^{D}.
\end{equation}
Here the power law refers to the geometric scaling of the correlation integral, i.e. how the cumulative number of contour points within distance \(r\) grows with \(r\), and it is unrelated to physical power laws such as turbulent energy spectra. The exponent \(D\) is the correlation dimension. 

For a simple line, the number of points increase linearly so that $D=1$, while for an area filling curve the number of points increases quadratically so that $D=2$. Thus, when applied to a line the increase in the number of points with radius is bounded from below and above by power laws of $r^1$ and $r^2$ (corresponding to $D=1$ and $2$), respectively. Values of $D$ between these extremes correspond to the line having some form of complexity. For true mathematical fractals, this complexity takes the form of a self-similar repeating pattern, and power laws with $1<D<2$ extend down to ever smaller scales. In practise however, in nature such complexity tends to be confined to a finite scale range \citep[e.g.][]{Vassilicos1991}, and it is in this sense that we apply this method here.

\begin{figure}[!t]
  \centering
  \includegraphics[width=0.95\columnwidth]{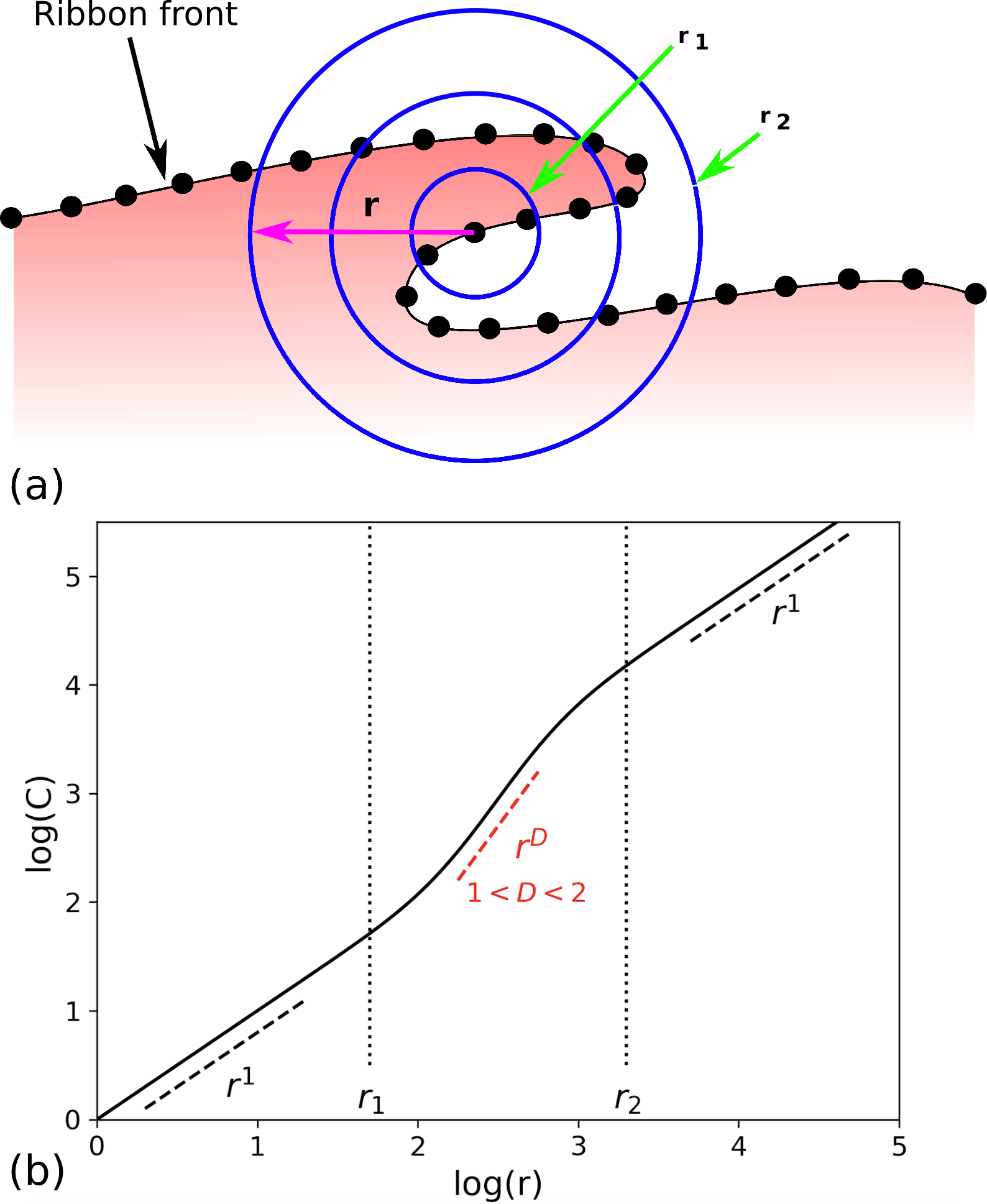}
  \caption{
  (a) Schematic showing circles with increasing radius within which the number of points on the ribbon front contour are counted as part of the correlation dimension calculation. (b) A schematic representation of the number of points ($C$) as a function of radius ($r$), see text for details.} 
  \label{fig:cartoon}
\end{figure}

In Figure \ref{fig:cartoon}(a) we display, in schematic form, how the correlation integral works in our case. For a point that is part of the ribbon front that forms the spiral (the point shown in the schematic), we take a set of concentric circles with increasing radius \(r\) and, for each \(r\), we add up the number  of points within \(r\). Figure  \ref{fig:cartoon}(b) shows a schematic representation of how $C(r)$ would be expected to increase with $r$. For small radii ($r<r_1$) the nearby front is a line, and $D\approx 1$. As $r$ increases beyond $r_1$, points brought closer due to the wrapping up of the front within the spiral begin to be counted, increasing the rate of change of $C(r)$. When plotted on a log-log plot, the gradient (and therefore $D$) increases above $1$, but will remain below $2$. Finally, once $r> r_2$ only new points outside of the spiral are counted and as the ribbon front is a line beyond this radius we once more find $D\approx 1$. 

The change in $C(r)$ with $r$ highlights that the length scale over which the CDM method is applied plays an important role in being able to identify these structures. Depending on the scale or internal structure of the spiral, the lower bound $r_1$ can extend to small scales but will ultimately be limited by the spatial resolution of the original image data. Whereas the upped bound corresponds to the maximum spatial extent of the spiral feature. In theory, the rate of change of $C(r)$ over the length scale of the spiral ($r_2 > r> r_1$) need not follow a power law itself. However, in practice (as discussed further in Sec. \ref{sec:results}) we assume this to be true by using a line of best fit to the log-log graph of $C(r)$ over length scales chosen to match those of a given spiral. 

When applied to all the points in the ribbon front, each point that forms a spiral will have this local increase in the gradient of $\log(C(r))$, but the precise value for $D$ and the length scale range over which the change occurs will be slightly different from point to point. However, collectively they will all have a value for the gradient in the steepest parts of their graphs greater than $1$ (and approximated by finding $D$) which we use as a practical way to identify points within locally wrapped up sections of ribbon front.

These points are subsequently processed through DBSCAN, a clustering algorithm sensitive to local point densities. DBSCAN groups points into clusters based on specified minimum density criteria, thus effectively isolating the spiral-like segments of the ribbon front. We use DBSCAN as a simple and transparent density-based clustering step because, in this simulation, the CDM-selected core points form
compact groups around spiral-/wave-like boundary segments and these groups are typically separated by long, smoother ribbon-front sections
with no core points. In this regime the candidate detections are generally spatially isolated, and DBSCAN robustly partitions them into
distinct clusters while rejecting sparse outliers. 
We refer the reader to Appendix~\ref{app:dbscan_hdbscan} for a justification of our choice of DBSCAN. We choose the DBSCAN neighbourhood scale
\(\epsilon\) from the inferred sampling/point spacing along the sampled contour (so that neighbouring core points within a few spacings are
grouped), while \texttt{min\_samples} sets the minimum local support required for a cluster to be retained. For completeness, we note that
hierarchical or adaptive density-based variants (e.g., HDBSCAN) can be advantageous when clusters have strongly varying densities or lie in close
proximity, exploring such alternatives is a natural extension for more challenging datasets, including some observational cases.

\subsection{Geometric characterisation of each spiral}\label{sec:ellipse-fit}
To provide a simple characterisation of each sub-structure, once each ribbon front 
feature is identified, it is fitted with a minimum-area ellipse using Welzl's algorithm \citep{Welzl1991}. This works as follows: let
$\mathcal{P}=\{\mathbf p_i=(\varphi_i,\theta_i)\in\mathbb R^{2}\}_{i=1}^{N}$
be the planar footprint of one of the clusters returned by the
CDM--DBSCAN method, where $\theta$ and $\phi$ denote the longitude and latitude coordinates (in degrees) in our simulation.   
Our aim is to monitor every ellipse/spiral with one simple curve that is
(1) easy to compare between frames and (2) insensitive to the
irregular sampling of the points.  
The natural choice is the minimum–area enclosing ellipse (MAEE),
i.e.\ the unique ellipse of smallest Euclidean area that contains all of
$\mathcal P$.
The MAEE problem reads
\begin{equation}
  \min_{\mathbf c,a,b,\theta}\;
        \pi a b
  \quad
  \text{subject to}\;
  \mathbf p_i\in \operatorname{ellipse}(\mathbf c,a,b,\theta)
  \ \ \forall i .
  \label{eq:maee-opt}
\end{equation}
\begin{figure*}[!h]
  \centering
  \vspace*{-50pt}
  \includegraphics[width=0.90\textwidth]{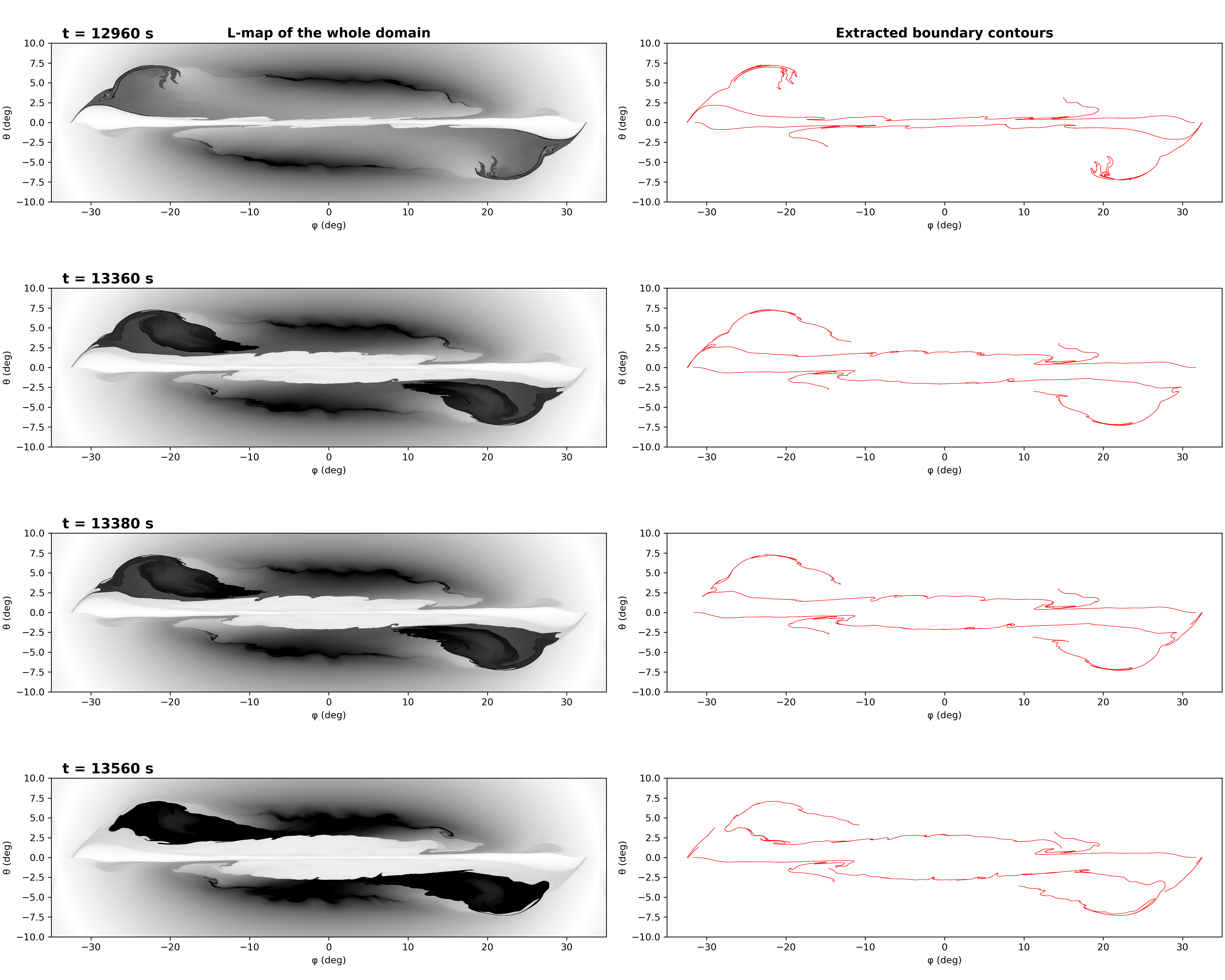}
  \vspace{-2mm}
  \caption{
    Evolution of the L-maps during the early stages of the
    simulated flare.  \textit{Left column:} $L$-maps of the full
    domain at four snapshots ($t = 12\,960$, 13\,360, 13\,380, and
    $13\,560\ \mathrm{s}$).  Darker tones correspond to longer traced
    field-line lengths, highlighting the erupting flux-rope footpoints.  \textit{Right column:} outer boundary contours
    automatically extracted from the same maps (Sec.~\ref{sec:contours}).  These
    contours form the basis on which the correlation-dimension and DBSCAN
    (Sec.~\ref{sec:data}) are applied to isolate spiral-like structures.}
  \label{fig:lmap_overview}
\end{figure*}
In this context, \(\mathbf{c}\) gives the coordinates of the ellipse centre, \(a\) and \(b\) are the semi-axis lengths (the semi-major and semi-minor axes, respectively), and \(\theta\) is the ellipse orientation angle.
In two dimensions this optimisation can be solved in expected linear time
in \(N\) by the randomised incremental algorithm of \citet{Welzl1991}.
We apply this algorithm to every cluster, the result is the parameter
vector \((\mathbf c,a,b,\theta)\), from which we compute \(\chi=a/b\).

This combined CDM-DBSCAN ellipse-fitting framework provides a comprehensive and automated approach to systematically detect the spirals that constitute the imprints of the plasmoids on the flare ribbons.
 
In the next section we will explain in detail how we apply CDM coupled with DBSCAN and MAEE and how well it performs.

\section{Application to the Lmap dataset}
\label{sec:application}
In the preceding section we developed our automated detection strategy in theoretical detail, here we demonstrate its application to a sequence of snapshots drawn from the 3-D MHD simulation.
\subsection{Method performance and application}
The workflow begins by tracing magnetic field lines at every snapshot and constructing the corresponding \(L\)-maps. The tracing is sampled on a \(5000 \times 4000\) grid with spacing $(\theta,\varphi) \in (0.004^\circ,0.0175^\circ)$. We choose a greater resolution in the $ \theta $ direction due to the elongation of the ribbons. Figure~\ref{fig:lmap_overview} shows four representative snapshots at 
\(t=12{,}960\), \(13{,}360\), \(13{,}380\), and \(13{,}560\;\text{s}\).
The earliest time (12 960 s) captures the onset of the flare, whereas the later times follow its progressive development.  This evolution is visible, for example, in the increasing separation of the two flare ribbons (the boundary between light grey and grey/black) a well-known signature of reconnection proceeding to ever-higher loops.
\subsubsection{Boundary--contour extraction}\label{sec:contours}
Starting from the field-line–length maps \(L(\varphi,\theta)\) computed for each
snapshot, we first rescale every map to the unit interval.  We then apply the
Canny edge detector \citep{Canny1986}.  The algorithm (i) smooths the image with
a derivative-of-Gaussian kernel to obtain both the gradient magnitude
\(\lvert\nabla\tilde L\rvert\) ( $\tilde L$ is the rescaled \(L(\varphi,\theta)\)) and its direction, (ii) performs
non-maximum suppression so that only single-pixel ridges aligned with the local
gradient remain, (iii) uses two intensity thresholds
to decide which pixels belong to an edge: pixels above \(T_{\mathrm{high}}\) are
kept as strong edge pixels, pixels below \(T_{\mathrm{low}}\) are rejected as
background, and pixels in between are kept only if they connect to a strong edge
through a continuous chain of neighbouring pixels (including diagonal
neighbours). Choosing adaptive limits
\(T_{\mathrm{low}} = 0.33\,m\) and \(T_{\mathrm{high}} = 1.33\,m\), where
\(m = \operatorname{median}(\tilde L)\), ties the sensitivity to the
instantaneous contrast of each map and removes the need for manual retuning. This retains faint but genuine edge segments while suppressing
isolated noise.
This yields a binary mask that traces
all steep gradients in the image.
Closed contours are next extracted with the Suzuki--Abe
border following algorithm \citep{SuzukiAbe1985}, implemented in
OpenCV as \texttt{cv2.findContours}.%
\footnote{In OpenCV the C library call
         \texttt{cvFindContours} was re-wrapped in \texttt{cv2};
         both ultimately invoke the Suzuki--Abe method.}

\begin{figure*}[!t]
\centering
\vspace*{-25pt}
  \includegraphics[width=1.4\columnwidth]{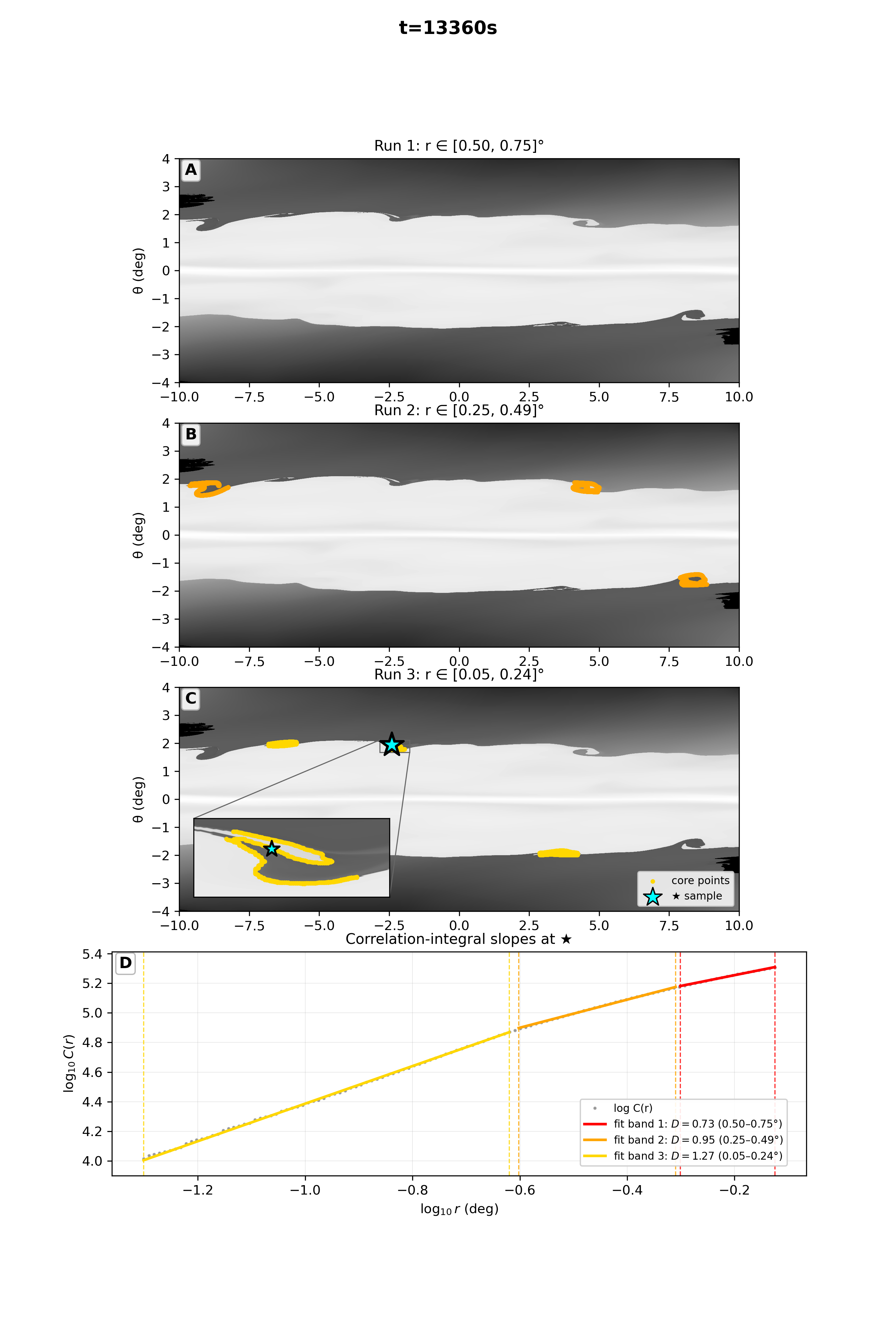}
  \caption{CDM results at \(t=13{,}360~\mathrm{s}\).
(A) Run 1, \(r\in[0.50^\circ,0.75^\circ]\): no spirals detected.
(B) Run 2, \(r\in[0.25^\circ,0.49^\circ]\): spiral segments identified (orange).
(C) Run 3, \(r\in[0.05^\circ,0.24^\circ]\): additional finer spirals, the star marks the sample used in (D). The inset shows a zoomed-in view of the detected spiral and the starred sample point. 
(D) \(\log_{10} C(r)\) versus \(\log_{10} r\) at the starred location with least-squares fits over the three bands, yielding slopes \(D=0.73\), \(0.95\), and \(1.27\), respectively.
}
  \label{fig:CDM_passes}
\end{figure*}

Because the native sampling density along a contour depends on the local
slope in pixel space, the extracted boundaries are re-sampled at uniform
arc-length before further analysis. For a given contour represented by a
vertex list $\{\mathbf p_i\}_{i=0}^{N-1}$ we compute the cumulative
arc-length $s_i = \sum_{j < i} \lVert \mathbf p_{j+1} - \mathbf p_j \rVert$ and
interpolate the curve at equidistant positions
$s_k = k\,h$ with step $h = 0.02^{\circ}$ using linear interpolation in the
$(\varphi,\theta)$ plane. The resulting point cloud
$\{\mathbf p_{\mathrm{new},k}\}$ possesses an approximately homogeneous
spatial density, a prerequisite for unbiased correlation-dimension estimates. 
As shown in the figure the method described above is able to preserve every kink and indentation, so the resulting contours capture the geometric detail along the two ribbon flare.

\begin{figure*}[!t]
\centering
\vspace*{-25pt}
  \includegraphics[width=0.90\textwidth]{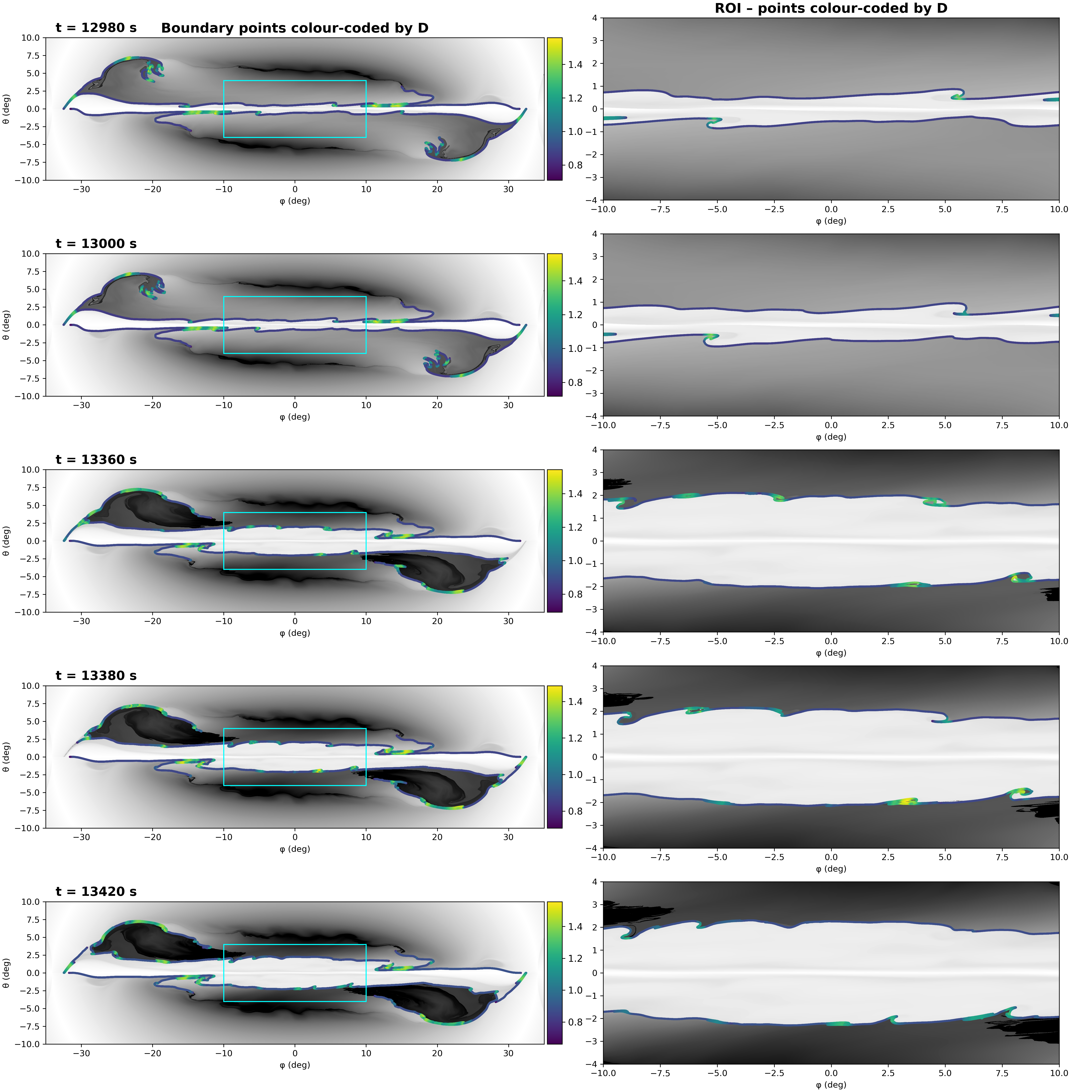}
  \vspace{-2mm}
  \caption{Local correlation-dimension \(D\) across five snapshots.
\textit{Left column:} full-domain \(L\)-maps with uniformly resampled boundary
contours, points are colour-coded by their estimated \(D\). The cyan rectangle marks the straight-section region of interest (ROI).
\textit{Right column:} zoom of the same ROI at the same times using the same
colour scale. Elevated values \(1<D<2\) highlight spiral-like segments along the ribbon boundaries, whereas values near \(D\!\approx\!1\) trace smoother ribbon edges. Times shown: \(t=\{12{,}980,\,13{,}000,\,13{,}360,\,13{,}380,\,13{,}420\}\,\mathrm{s}\).}
  \label{fig:spiral_identification}
\end{figure*}
\subsection{3-run CDM and the performance of the method}
We evaluate the correlation dimension in three successive bands of radii \(r\) so that each pass targets a different spatial scale of structures
along the ribbon boundary. Specifically,
\[
\begin{aligned}
\text{Pass~1:} &\quad r\in[0.50^\circ,\,0.75^\circ],\\
\text{Pass~2:} &\quad r\in[0.25^\circ,\,0.49^\circ],\\
\text{Pass~3:} &\quad r\in[0.01^\circ,\,0.24^\circ].
\end{aligned}
\]
Within each pass we compute the local correlation integral and fit \(C(r)\propto r^{D}\); samples with \(D>1\) are kept as core spiral candidates and clustered with \textsc{DBSCAN}. The survivors from a pass are excluded from the next one, so the procedure proceeds from coarse to fine scales without duplicating detections. The wide (Pass~1) band suppresses pixel-scale noise and only responds to very broad, high-contrast spirals, the mid band (Pass~2) captures the typical
spacings of most spirals in this dataset, and the fine band (Pass~3) recovers narrow, spiral segments that would be washed out at larger \(r\). The estimate for $D$ at a given point is therefore given by the first value it attains above the threshold across the 3 scans, with points below the threshold taking the value found in the final scan. This multiscale, coarse-to-fine strategy raises sensitivity across the full range of observed spiral widths while controlling false positives and keeping the detections non-overlapping. 

We demonstrate how this approach performs in Figure \ref{fig:CDM_passes}.
Panels A–C show points identified within the straight sections of the two ribbons in contrast against the grayscale \(L\)-map background. 
Orange points mark locations where the local CDM fit returns \(D>1\): Pass~1 (A) finds none for this snapshot, Pass~2 (B) reveals several mid-scale spirals, and Pass~3 (C) adds finer spirals. The cyan star marks a representative point used for the slope analysis. Panel D plots \(\log_{10} C(r)\) versus \(\log_{10} r\), dashed vertical lines delimit the three radius bands and straight-line fits within each band yield the local dimensions (here \(D\simeq0.73,\,0.95,\,1.27\)). Values near \(D\!\approx\!1\) correspond to smooth ribbon edges, whereas elevated \(1<D<2\) indicate spiral-like clustering at the corresponding scales. The estimate \(D<1\) arises from geometry and finite windowing of the correlation integral. The correlation integral counts neighbours within a circle of radius \(r\), along nearly straight segments, each circle intersects less arc length than on curved parts, so \(C(r)\) grows sub-linearly over the fitted radii and the log–log slope falls below one. This is amplified when the regression window approaches the available scale limit (\(r\to r_{\max}\)), where \(C(r)\) naturally flattens. Crucially, our analysis uses \(D\) as a relative complexity measure rather than as an absolute fractal dimension.

\begin{figure*}[!t]
  \centering
  \includegraphics[width=\textwidth]{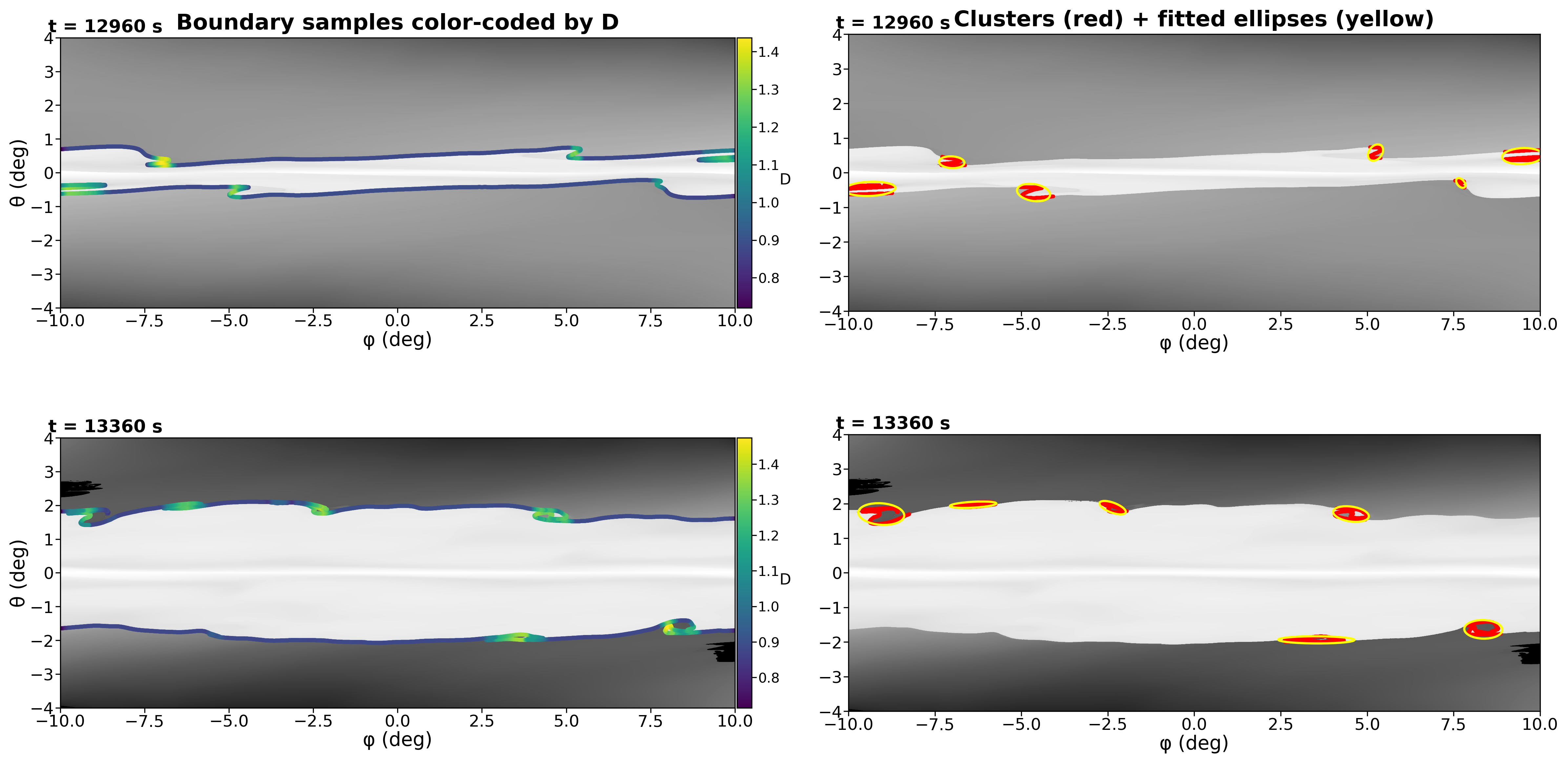}
  \vspace*{-10pt}
  \caption{CDM--DBSCAN detections at two representative snapshots. Left panels: uniformly sampled ribbon boundary overlaid on the grayscale $L$-map and colour-coded by the local correlation dimension $D$, where elevated $D$ highlights spiral-like structure. Right panels: points with $D>1$ that are grouped by DBSCAN are shown in red, with a minimum-area enclosing ellipse (yellow) fitted to each cluster. Rows correspond to $t=12960$ s (top) and $t=13360$ s (bottom). The accompanying animation shows the same layout for the full time sequence, illustrating that spiral detections appear intermittently at early times, become more numerous during the more developed phase of the flare and then diminish towards the end of the event.}

  \label{fig:performance}
\end{figure*}

\begin{figure}[t!]
\hspace*{-0.1\columnwidth}
  \includegraphics[width=1.1\columnwidth]{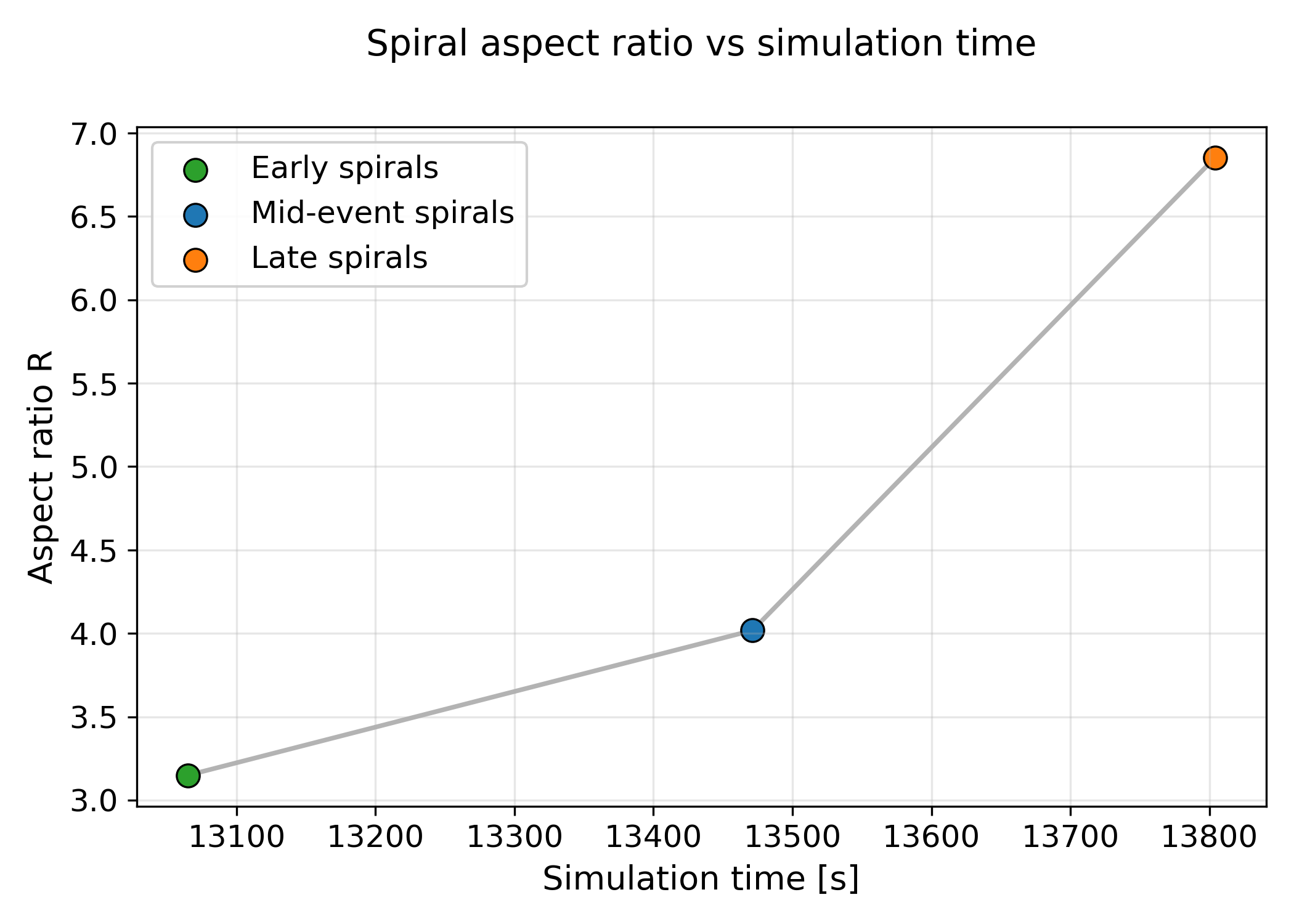}
  \caption{Mean spiral aspect ratio versus simulation time. Each coloured point shows the mean ellipse aspect ratio $R = a/b$ for spirals in three temporal cohorts: an early phase (green; spirals starting before $t=13200$~s), a mid-event phase (blue; $13200\le t_{\mathrm{start}}\le 13650$~s), and a late phase (orange; $t_{\mathrm{start}}>13650$~s). Points are plotted at the mean midpoint time of the spirals in each cohort, and the grey line connects these group means. }

\label{fig:mean_aspect}
\end{figure}

Figure~\ref{fig:spiral_identification} illustrates the performance of the \(D\) evaluation. The left column shows L-maps of the two-ribbon setup at times \(t=\ 12\,980,\,13\,000,\,13\,360,\,13\,380,\,13\,420\\\mathrm{s}\), with sampled boundary points overplotted and coloured by their estimated correlation dimension \(D\). The cyan rectangle marks the straight section of the ribbons, the region we focus on hereafter. 
The right column displays zooms of the marked region at the same times. Overall, these plots show that the three-pass CDM approach highlights spiral-like structures; points within spirals show elevated \(D\), making them stand out from the surrounding ribbon. We highlight that during the early phase of the flare \(\bigl(13\,360\text{--}13\,420~\mathrm{s}\bigr)\), the straight section of the ribbons hosts fewer spirals than in the later snapshots \(\bigl(12\,980\text{--}13\,000~\mathrm{s}\bigr)\), consistent with an increasing rate of plasmoid formation as the eruption progresses.
The surviving points of a given pass define a “core’’ set that is clustered with DBSCAN \(\bigl(\varepsilon=0.06^{\circ},\ \textit{min\_samples}=5\bigr)\), isolating coherent spiral segments along the extracted contours. Here, \(\varepsilon\) is the neighborhood radius in the chosen clustering metric specifically, the maximum Euclidean separation in the \((\theta,\phi)\) angular plane for two points to be considered neighbours so \(\varepsilon=0.06^{\circ}\) defines the radius of each \(\varepsilon\)-ball. Two points in the \((\phi,\theta)\) plane are considered neighbours if their separation is at most \(\varepsilon\) and \(\textit{min\_samples}\) is the minimum number of points within an \(\varepsilon\)-ball (including the point itself) required for a point to be labelled a core point. Clusters are formed by connected core points, points lying within \(\varepsilon\) of any core are treated as border points, and all remaining points are designated as noise. The minimum-area enclosing ellipse algorithm is then applied to each CDM cluster, fitting an ellipse that summarises its geometry.
In Figure~\ref{fig:performance} we summarise the performance of our method by showing two representative snapshots at \(t=12\,960~\mathrm{s}\) (early phase) and \(t=13\,360~\mathrm{s}\) (more evolved phase).
The left column shows \(L\)-maps with sampled boundary points overplotted and coloured by their estimated correlation dimension \(D\), as in Figure~\ref{fig:spiral_identification}. The right column displays the detected spirals: DBSCAN clusters of points with \(1<D_i<2\) (red), together with the corresponding ellipses fitted using the MAEE algorithm (yellow contours). 

Overall, the method yields consistent and robust detections, automatically isolating the
spiral-like segments of the boundary contours within the ribbon’s straight section. Because
the outcome is sensitive to the threshold choices in both the CDM and the DBSCAN clustering,
we recommend a dedicated parameter scan before the method is deployed on a new data set.

The fitted ellipses allow us to automatically track spirals in time, and the fully automated
workflow enables analysis of the numerous spirals present in our high-resolution 3-D MHD simulation. 
For a new data set, a sensible starting point is to choose the CDM scale bands from the visible size of the spiral-like features and the effective resolution of the map. As a rule of thumb, set the upper radius bound $r_{\max}$ to be of the same order as the largest coherent spiral-like segments you can identify by eye (in our simulation these are $\sim 1^\circ$, which motivated the largest $r$ tested), and set the lower bound $r_{\min}$ to be a few resolution elements so that the fit is not dominated by pixel-scale noise. With these bounds fixed, the intermediate bands can be spaced logarithmically between $r_{\min}$ and $r_{\max}$, and the D-threshold can then be refined in small steps while monitoring a few simple diagnostics: (i) visually overlaying the detected spirals on the $L$-maps to check that they trace coherent spiral-like segments rather than isolated noise, (ii) checking that the number and location of detected spirals change smoothly when the D-threshold or the DBSCAN parameters are varied by $10$–$20\%$, rather than jumping catastrophically, and (iii) verifying that only a limited fraction of the ribbon boundary is flagged as “spiral-like”; too few detections suggests an overly strict threshold, while complex segments covering most of the ribbon suggests an overly permissive one. If these checks are satisfied, our experience is that the resulting spiral catalogue and the derived statistics are robust to the exact parameter choices, even though the fine details of the masks do depend on the particular data set.

\begin{figure*}[!t]
  \centering
  \includegraphics[width=\textwidth]{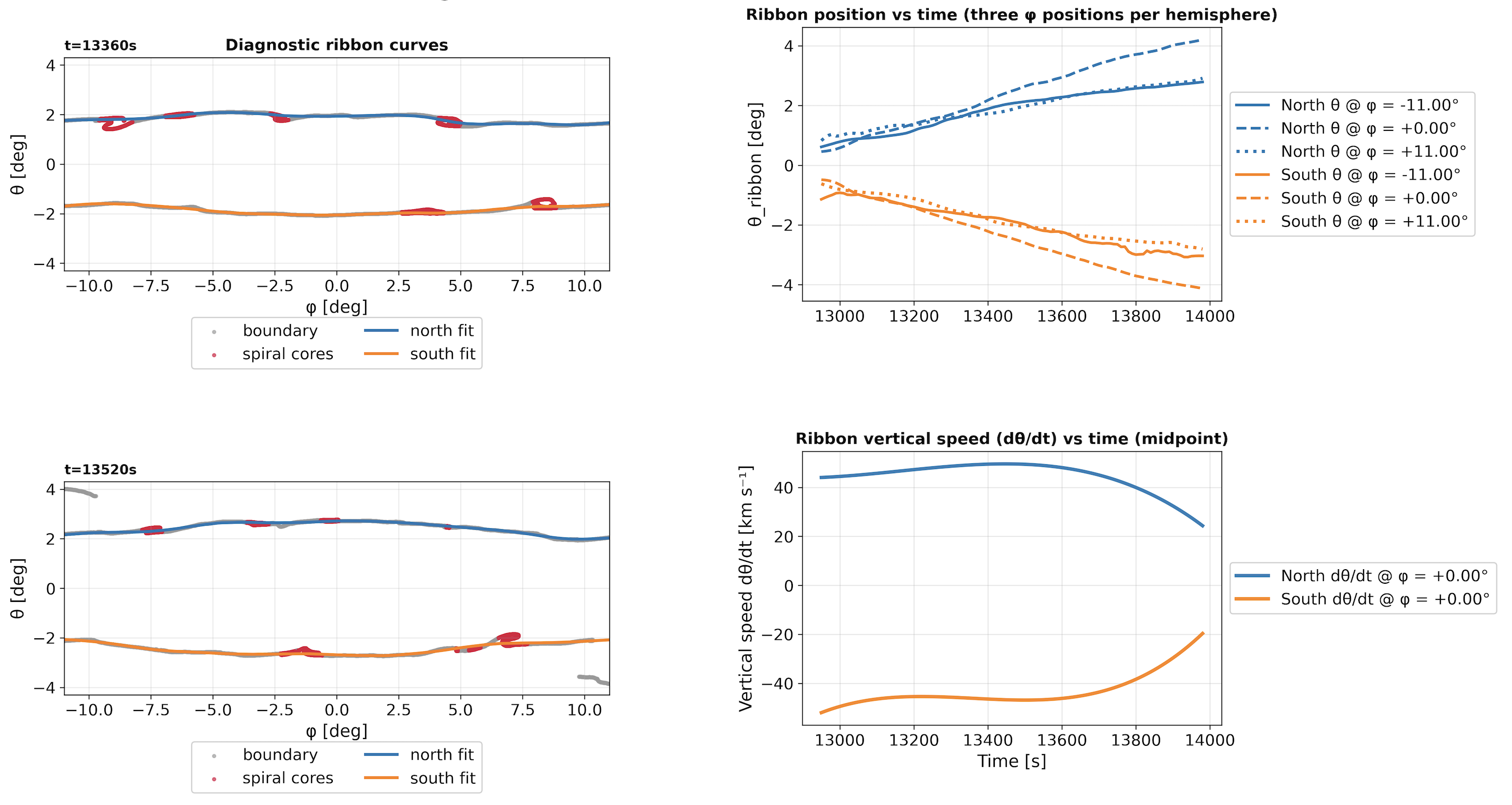}
  \vspace*{-10pt}
  \caption{Ribbon average curve and kinematics. \textbf{Left:} snapshots at
  $t=13360\,\mathrm{s}$ and $t=13520\,\mathrm{s}$; grey points are boundary samples,
  red points are spiral points (excluded), and blue/orange curves are the
  fitted north/south lines across the straight section.
  evolution of $\theta_{\rm ribbon}(t)$ at three longitudes per hemisphere
  shows coherent separation. 
  \textbf{Bottom right:} normal speed of the midpoints}.
  \label{fig:true_ribbon_curve}
\end{figure*}
\begin{figure*}[t]
  \centering
  \includegraphics[height=.50\textheight]{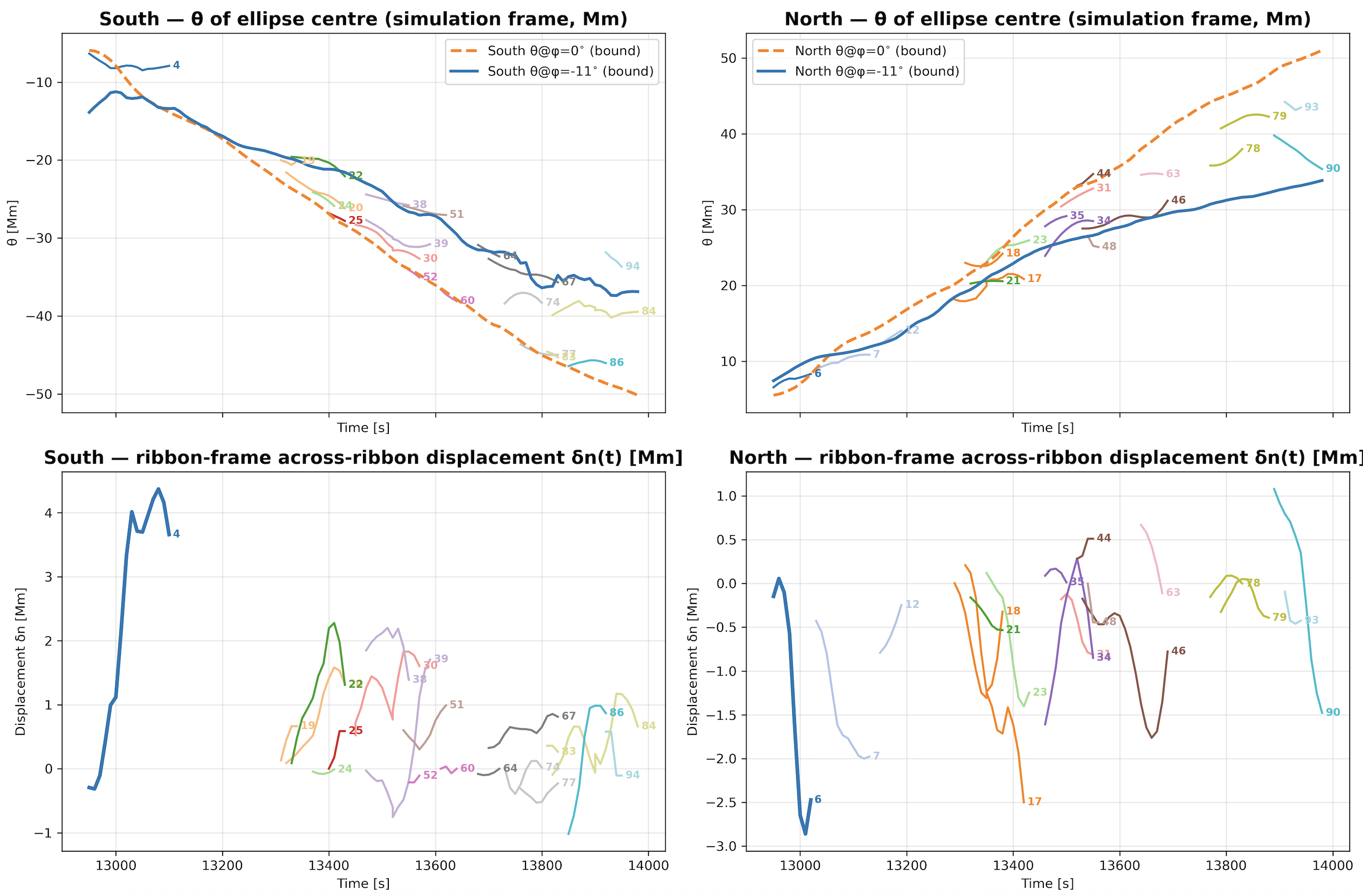}
  \vspace{-1mm}
  \caption{Latitudinal motion and across–ribbon displacement of detected spirals. 
\textit{Top row:} latitude $\theta(t)$ of each ellipse centre (thin coloured curves, labels are spiral IDs) in the fixed simulation frame for the south (left) and north (right) ribbons. The overplotted references are the fitted ribbon tracks: dashed line is the ribbon latitude at the midpoint longitude $\theta_{\rm ribbon}(t)\big|_{\phi=\mathrm{mid}}$, and solid line is the latitude at the western edge $\theta_{\rm ribbon}(t)\big|_{\phi=\min}$.
\textit{Bottom row:} ribbon–frame normal (across–ribbon) displacement $\delta n(t)$ in Mm for the same spirals. }

  \label{fig:theta_dual}
\end{figure*}
\begin{figure*}[t]
  \centering
  \includegraphics[height=.50\textheight]{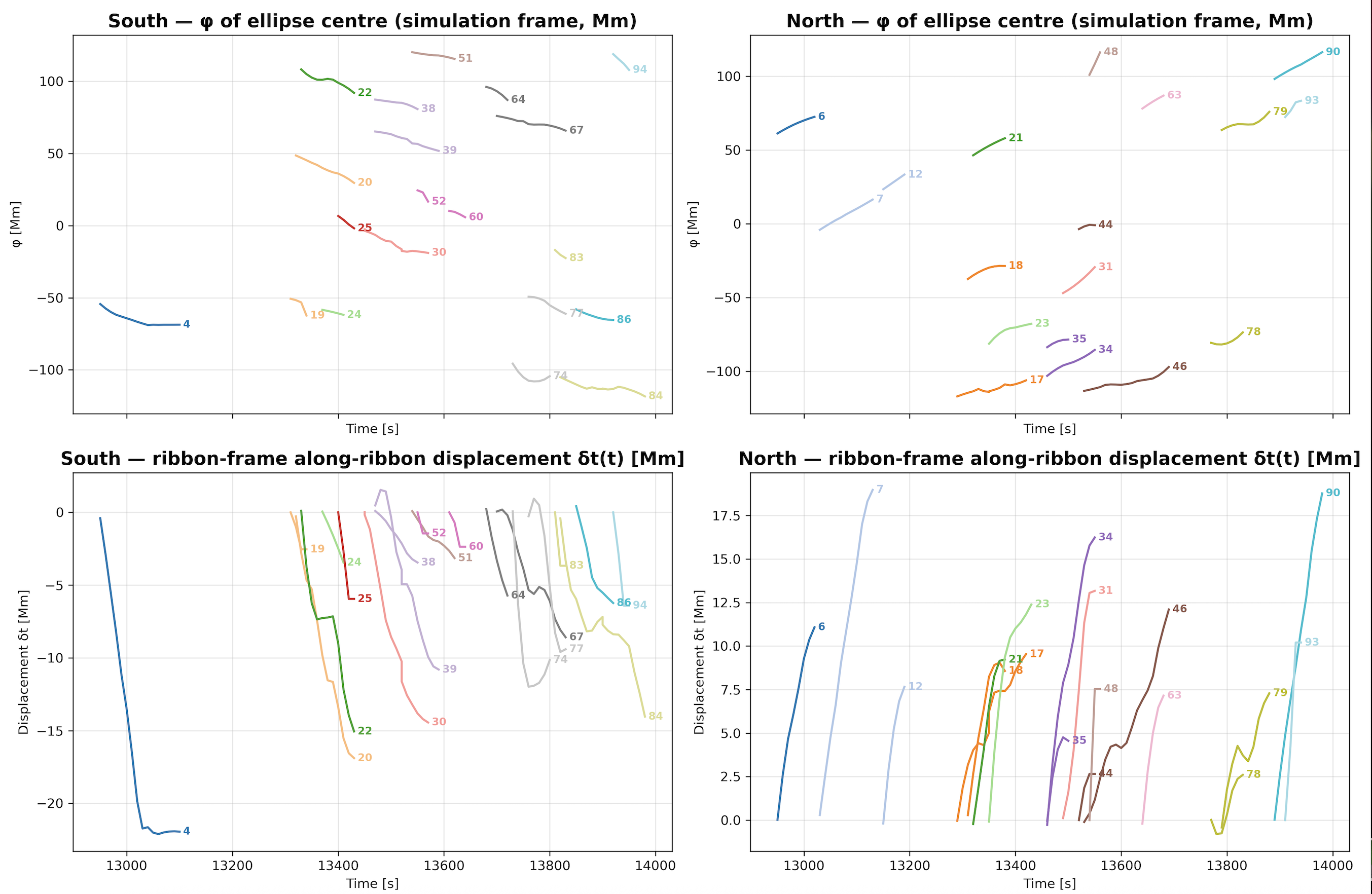}
  \vspace{-1mm}
  \caption{Along–ribbon motion of spiral centres by hemisphere.
\textit{Top row:} longitude $\phi(t)$ of each ellipse centre in the fixed
simulation frame for the south (left) and north (right) ribbons; thin coloured
curves are individual tracks (labels are spiral IDs).
\textit{Bottom row:} ribbon–frame along–ribbon displacement $\delta t(t)$ in\,Mm,
measured as the arclength change of each centre projected onto the fitted
ribbon, relative to its first detection.}
  \label{fig:phi_dual}
\end{figure*}

\begin{figure*}[t]
  \centering
  \includegraphics[height=.50\textheight]{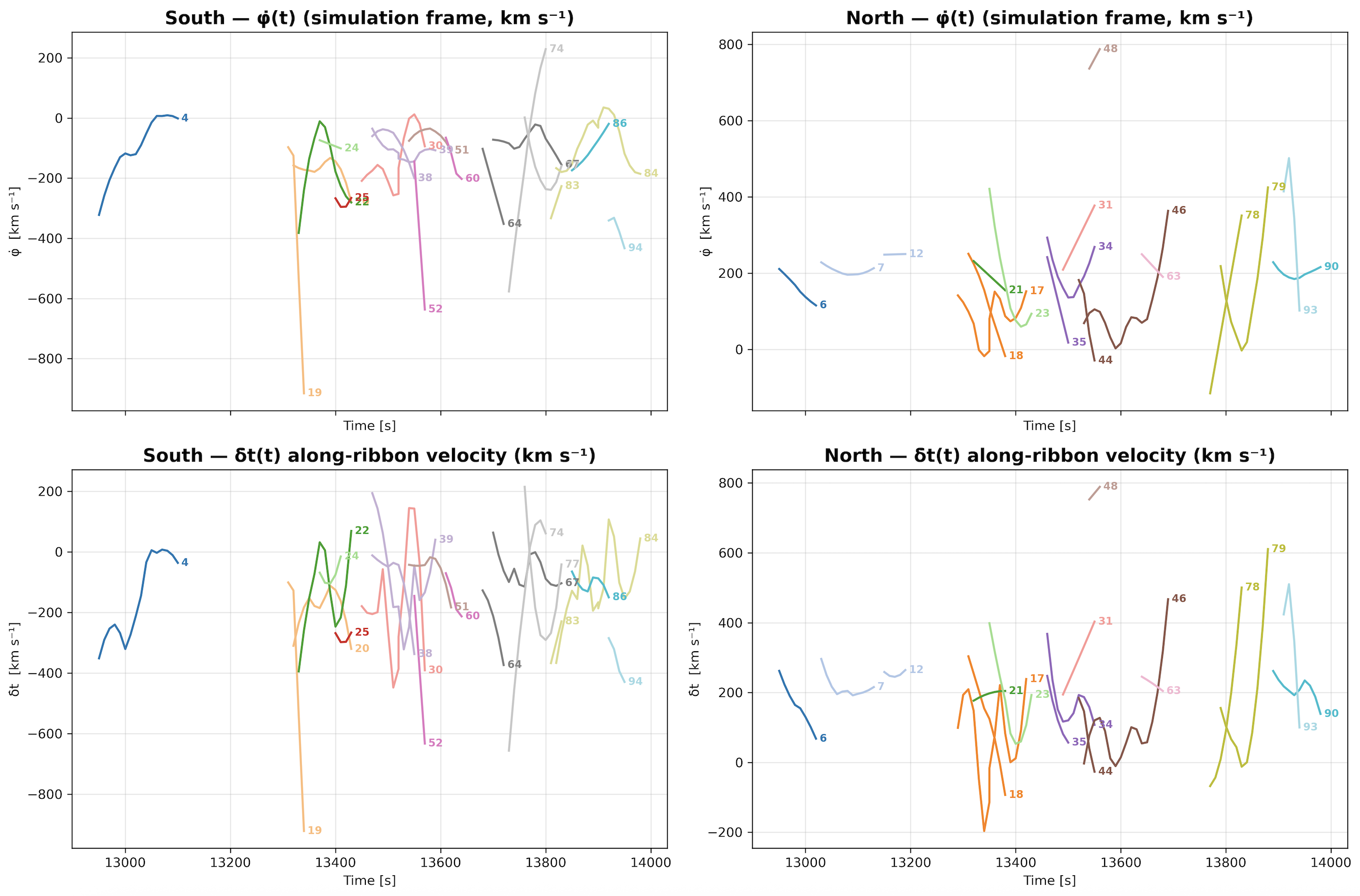}
  \vspace{-1mm}
  \caption{Along–ribbon drift speeds of spiral centres.
\textit{Top row:} instantaneous longitudinal speed $\dot\phi(t)$ of each ellipse
centre in the simulation frame, reported in km\,s$^{-1}$, thin coloured curves
are individual spirals.
\textit{Bottom row:} corresponding along–ribbon speed in the ribbon frame,
$\dot{\delta t}(t)$, obtained by differentiating the arclength coordinate of
each centre projected onto the fitted ribbon.}
  \label{fig:phi_dual_velocity}
\end{figure*}

\section{Statistical analysis of the dataset}\label{sec:results}

\subsection{Overview}
Figure \ref{fig:performance} and the accompanying movie demonstrates that our method robustly detects spirals along the straight sections of the flare ribbons throughout the flare's evolution. Furthermore, across the full dataset we observe a clear sequence: an initial, short–lived burst of rapidly moving plasmoids, a brief gap, then a more developed, bursty/turbulent reconnection interval, and finally a gradual waning phase in which spirals diminish in size and occurrence. We interpret the first burst as the onset of reconnection that builds the main flux rope and ribbon, with the subsequent interval reflecting a later, more developed and intermittently turbulent reconnection regime. 
 This evolution coincides with a progressive decline in the guide–field strength \citep{Joel_2022aApJ...932...94D}. Kinematically, the detected spirals drift coherently away from the ribbon hooks, and their perpendicular motion closely tracks the mean outward advance of the ribbon. In the movie the spirals first appear only sporadically before $t \approx 13200$ s. Their occurrence rate then increases sharply, with many spirals detected in short bursts up to $t \approx 13650$ s, before declining again towards the end of the flare. They also appear to evolve geometrically, from nearly circular footprints to more elongated shapes as the  flare progresses. To quantify this, we computed the mean ellipse aspect ratio $R = a/b$ for each tracked spiral throughout its lifetime and then averaged $R$ over three temporal cohorts: an early group (spirals with start times before 13200 s), a mid-event group (start times between 13200 and 13650 s), and a late group (start times after 13650 s). The resulting trend (Fig.~\ref{fig:mean_aspect}) shows a systematic increase in $R$ with simulation time, from $R \approx 3$ in the early group to $R \approx 7$ late in the event, confirming that the spirals become progressively more elongated. As will be discussed further in Sect. 5.4, the time periods that define the groupings used above correspond roughly with what would be termed the onset, impulsive and decay phases of the flare \citep[e.g.][]{Fletcher2011SSRv,Benz2017LRSP} in this simulation.

\subsection{Sub-structure velocities}
The key advantage of our method is its fully automatic operation, which enables large-sample
statistical analysis of the numerous spirals present in the 3-D numerical simulation. This, in turn, will allow us in the future to investigate and quantify links between these spirals and the dynamic behaviour of plasmoids in the flare current sheet.
To conduct an accurate kinematic analysis of the ribbon features, the background motion of the ribbon front must be quantified and accounted for. To do so, we consider only points within the reduced region of interest ($\varphi \in \pm 11^\circ$).
We then remove the points assigned to spiral  
clusters from the overall set of points defining the boundary contours extracted from the L-maps 
within this region. Finally, we then reconstruct the ribbon's average curve 
$\theta_{\mathrm{ribbon}}(\phi,t)$ for each ribbon front using LOESS (\textit{LO}cally \textit{E}stimated \textit{S}catterplot \textit{S}moothing)
\footnote{also called LOWESS: LOcally \textit{W}eighted \textit{S}catterplot \textit{S}moothing} by fitting a  weighted polynomial around each target longitude $\phi_0$ on a uniform evaluation grid spanning the display range (the grid is only where we evaluate, the fit always uses the actual boundary samples $\{(\phi_i,\theta_i)\}$ at time $t$). 
For each $\phi_0$ we choose a local neighborhood of $k=\max\{\lceil fN\rceil,k_{\min}\}$ nearest data points by distance $|\phi_i-\phi_0|$ (with fraction $f\!\approx\!0.22$, total samples $N$, and a small floor $k_{\min}$ for stability), set $r$ to the distance from $\phi_0$ to the $k$-th neighbor, and assign tricube weights $w_i(\phi_0)=\bigl(1-(|\phi_i-\phi_0|/r)^3\bigr)^3$ for $|\phi_i-\phi_0|<r$ and $0$ otherwise so the closest points dominate. We then solve the weighted least-squares problem \begin{equation}
\min_{\beta_0,\beta_1,\beta_2}\sum_i w_i(\phi_0)\,\bigl[\theta_i-\beta_0-\beta_1(\phi_i-\phi_0)-\beta_2(\phi_i-\phi_0)^2\bigr]^2,
\end{equation} where $\theta_i$ and $\phi_i$ are the observed latitude/longitude of sample $i$, $w_i(\phi_0)$ are the kernel weights at $\phi_0$, and $(\beta_0,\beta_1,\beta_2)$ are the local quadratic coefficients (intercept, slope, curvature) centered at $\phi_0$, the fitted ribbon value is the intercept $\widehat{\theta}_{\mathrm{ribbon}}(\phi_0,t)=\widehat{\beta}_0$, and the longitudinal slope can be taken as $\partial\theta/\partial\phi(\phi_0,t)\approx\widehat{\beta}_1$ or computed by differentiating the evaluated grid. This sliding, locally weighted fit yields a curve that closely follows the boundary geometry. 

Figure~\ref{fig:true_ribbon_curve} brings together the curve we fit to each ribbon and the kinematics derived from that fit. The two panels on the left are diagnostic snapshots at $t=13520$\,s (top) and $t=13360$\,s (bottom): grey points mark the raw boundary sampled points, red points identify spiral cores that are excluded from the fit in order to focus on the average
ribbon curve, and the blue/orange curves are the north/south fitted lines for each straight section of ribbon front. 
The fitted lines track the bulk ribbon edges cleanly and remain stable: in these two times we see the expected outward displacement, northward for the northern ribbon and southward for the southern, consistent with progressive flare reconnection.
The top‐right panel then follows the ribbon latitude $\theta_{\rm ribbon}(t)$ at three representative longitudes, at one end of the fitted line, in the middle and at the other end (solid, dashed, dotted) for each hemisphere.  Both hemispheres show that the midpoint longitude exhibits the largest displacement in latitude ($\theta$) relative to the left and right reference longitudes. This behaviour in the L-maps is linked with the rate at which the magnetic connectivity changes perpendicular to the surface in the current sheet volume, since the surface magnetic flux varies little in $\varphi$ across this region this implies that reconnection proceeds faster near the centre of the straight section, close to $(\phi,\theta)\approx(0,0)$, than toward its ends.

The bottom‐right panel shows the separation speed at the midpoint longitude, obtained from the time derivative of $\theta_{\rm ribbon}$, $v=\partial\theta/\partial t$ (converted to km\,s$^{-1}$). By our sign convention, positive values denote motion toward $+\theta$. Over the interval shown the northern ribbon moves at $\sim45$~km\,s$^{-1}$ and the southern at comparable magnitude but opposite sign, with both exhibiting a gentle deceleration toward the end. Together, these panels demonstrate that a single differentiable fitted line provides a robust moving reference for the flare ribbon: it captures the bulk ribbon separation against which we can do a careful
kinetic analysis of the detected spirals.

\begin{figure}[t]
  \centering
  \includegraphics[width=\linewidth]{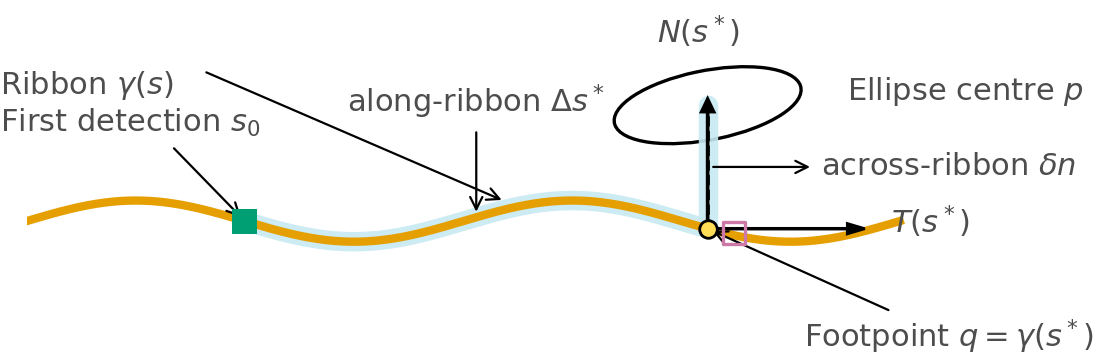}
  \caption{Schematic of the ribbon–frame kinematics. The fitted ribbon is a smooth curve $\gamma(s)$ in $(\phi,\theta)$. 
  For an ellipse centre $p=(\phi_c,\theta_c)$, the orthogonal projection onto the ribbon is $q=\gamma(s^*)$. 
  The unit tangent $T(s^*)$ and unit normal $N(s^*)$ define the local ribbon frame. 
  The across–ribbon displacement is $\delta n=(p-q)\!\cdot N(s^*)$, while the along–ribbon coordinate is the arclength $s^*(t)$; 
  we report the along–ribbon drift relative to first detection as $s^*(t)-s^*(t_0)$.  
  } 
  \label{fig:ribbon_cartoon}
\end{figure}

We analyze the motion of the detected spirals by tracking the evolution of the center of each fitted ellipse in both latitude (\(\theta\)) and longitude (\(\phi\)). To disentangle the intrinsic drift of the spirals from the bulk motion of the ribbons, we evaluate each center in two coordinate frames: (i) the simulation (global) frame, using the raw \((\phi,\theta)\) coordinates from the numerical domain, and (ii) the ribbon frame. For the ribbon frame and for each time frame and hemisphere, we take the fitted ribbon  as a smooth planar curve
\(\gamma(s)=(\phi(s),\theta(s))\) in the \((\phi,\theta)\) plane, parametrized by arclength \(s\).
Given an ellipse center \(p=(\phi_c,\theta_c)\), we compute its orthogonal projection onto the ribbon,
\(q=\gamma(s^*)\), where \(s^*\) is the arclength position of the closest point on \(\gamma\)
(implemented via a 
nearest-point search on the fitted line).
At \(q\) we evaluate the unit tangent and unit normal,
\[
T(s^*) \;=\; \frac{d\gamma}{ds}(s^*) \;/\; \Bigl\lVert \tfrac{d\gamma}{ds}(s^*) \Bigr\rVert,
\qquad
N(s^*) \;=\; R_{+90^\circ}\,T(s^*),
\]
with \(R_{+90^\circ}\) denoting a \(+90^\circ\) rotation in the \((\phi,\theta)\) plane while N is the unit normal and T is the unit tangent.
The across-ribbon displacement is the signed normal offset
\[
\delta n(t) \;=\; \bigl(p-q\bigr)\cdot N(s^*),
\]
and the along-ribbon coordinate is the arclength position \(s^*(t)\) of the footpoint.
We report the along-ribbon drift 
relative to the first time that a spiral is detected \(t_0\) as
\[
\delta t(t) \;=\; s^*(t)-s^*(t_0).
\]
We display the schematics of the ribbon-frame kinematics in Figure \ref{fig:ribbon_cartoon}. 

In Figure~\ref{fig:theta_dual} we show the latitudinal ($\theta$) displacement of the centre of each fitted ellipse in two frames: the fixed simulation frame (top row) and the ribbon frame (bottom row). In the top row we overlay the time-distance tracks of the fitted average ribbon curve at the midpoint and at the western edge of the straight section (the same curves as in Figure~\ref{fig:true_ribbon_curve}). For visual clarity, spirals with lifetimes shorter than $30\,\mathrm{s}$ are omitted (the cadence of the simulation
magnetic field data is 10s) . Across the straight section, all detected spirals move vertically in step with the average ribbon curve. This behaviour is seen more clearly in the bottom row, which plots each spiral’s across–ribbon (normal) displacement in the ribbon frame: for all detected spirals these displacements are small (compared to the overall displacement of the ribbon over the same time period), indicating that the spirals are kinematically locked to the ribbon and therefore follow its vertical motion. We highlight that the early spirals (4 and 6) have greater across-ribbon displacement in comparison with the spirals in the subsequent phases. This indicates the spirals in the onset phase have a slightly different behaviour than in later phases, suggesting a possible link to the guide field. 
However, to examine this further, it is essential to have more data/spirals in the onset phase which this dataset lacks.  Computing displacements in the ribbon frame is essential and makes our inference about the spirals' vertical motion robust.

In Figure~\ref{fig:phi_dual} we present the longitudinal ($\phi$) displacement of the centre of each fitted ellipse in the same two frames as Figure~\ref{fig:theta_dual}: the fixed simulation frame (top) and the ribbon frame (bottom). In the top row, all spirals detected in the straight sections of both ribbons exhibit clear lateral drift along the ribbon. Similar along–ribbon motions have been discussed in theory and observations (e.g., theory: \citealp{Priest1995JGR...10023443P,Aulanier2012A&A...543A.110A,Janvier2013A&A...555A..77J,Wyper_Pontin2021ApJ...920..102W}, observations: \citealp{Dudik2014ApJ,LiZhang2015ApJL,Dudik2016ApJ,Lorincik2019ApJ}), but here we quantify them systematically across a large data set using automatic detection.
The drift direction is opposite in the two ribbons: in the south ribbon the spirals migrate towards negative longitudes ($-\phi$), whereas in the north ribbon they migrate towards positive longitudes ($+\phi$). In both ribbons the spirals move away from their respective hooks, consistent with the analytical model of \citet{Wyper_Pontin2021ApJ...920..102W}. The ribbon–frame representation (bottom row), which plots the along–ribbon displacement $\delta t(t)$, sharpens these trends: within each ribbon the spirals are drifting in the same direction, confirming a coherent, unidirectional along–ribbon propagation shared by all spirals in a given ribbon. We observe that the magnitude of the displacement is the same in both the ribbons so there is not any geometrical preference of the lateral drift.

\par Next, we assess the drift velocity of the spirals. 
Instantaneous speeds are obtained from central finite differences of the tracked positions in time, and the resulting profiles are shown in Figure~\ref{fig:phi_dual_velocity}. Although the two reference frames yield slightly different details, the inferred speeds are broadly consistent in the range of values obtained, falling between \(10\)--\(800~\mathrm{km\,s^{-1}}\)
These motions are \textbf{sub-Alfv\'enic} with respect to the surface Alfvén speed: from the simulation we find a local Alfv\'en speed of \(v_A \approx 3000~\mathrm{km\,s^{-1}}\), the drifts correspond to \(\sim 0.003\)--\(0.27\,v_A\).
For context, along–ribbon motions during slipping reconnection can approach or even in theory exceed \(v_A\) (e.g., \citealp{LiZhang2015ApJL,Dudik2016ApJ,Lorincik2019ApJ,Lorincik2025NatAs...9...45L}); {with} lateral drifts exceeding \(1000~\mathrm{km\,s^{-1}}\) have also been reported \citep{Lorincik2025NatAs...9...45L}. Our measured speeds are therefore well below those extremes, an important point for observers.
This disparity suggests a different origin: we interpret the observed drift as footpoint motion driven by gradual, local changes in magnetic connectivity associated with plasmoid dynamics, rather than classical slip–running reconnection.
We also note a systematic increase in speed after \(t \gtrsim 13{,}400~\mathrm{s}\). A plausible explanation is that, as the flare evolves and the plasmoid/HFT topology becomes more intricate, the mapping gradients sharpen and split, leading to larger apparent footpoint displacements for the same underlying field-line evolution, and hence higher measured drift speeds. Nonetheless, these trends could be linked with the 3D motion of the plasmoids in the volume and their rapid changes in their magnetic connectivity. A detailed analysis of these trends and their connection to 3-D reconnection physics lies beyond the scope of the present paper, which focuses on introducing the automatic method and demonstrating its robustness, and will be pursued in future work.

\subsection{Magnetic flux}
By utilising the surface flux within the fitted ellipses, we can also obtain an estimate of the magnetic flux threading the plasmoids associated with the spiral ribbon sub-structures. For clarity, ``flux'' denotes the normal magnetic flux computed on the surface defined by the fitted ellipse of each detected spiral.

We note that as shown in \citet{Dahlin2025} particular plasmoids map to multiple spiral features on the surface, with the evolution of the proportion of flux threading a plasmoid connecting to a given spiral depending upon the dynamics of the plasmoid in the current sheet. As such, the amount of flux within a given spiral is likely a lower bound for the flux within a given associated plasmoid. Nevertheless, from a statistical point of view the distribution of flux may well follow similar trends to that of the true features within the current layer. Because comparable studies for fully 3D plasmoids are scarce, however, we proceed with appropriate caution. We test for a possible power-law relationship between spiral flux and spiral counts as similar power-law relationships have been inferred from 2D studies \citep[e.g.][]{PoP2012_LoureiroUzdenskyConfirm}.

\begin{figure}[t!]
  \includegraphics[width=1.0\columnwidth]{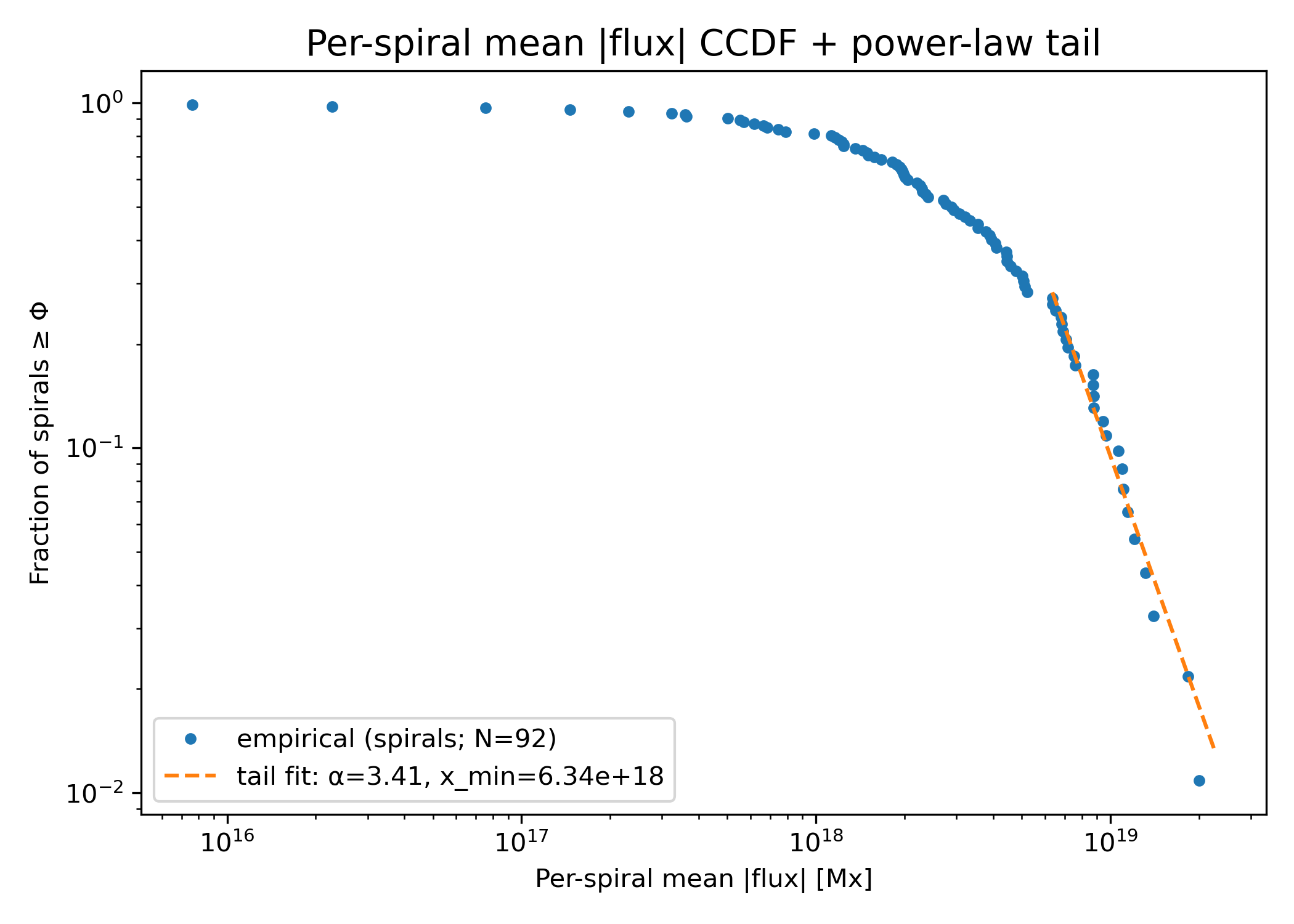}
  \caption{Complementary cumulative distribution of per-spiral mean unsigned flux.
  Blue points: empirical CCDF $S(\Phi)=\Pr(\Phi'\!\ge\!\Phi)$ for $N=92$ spirals,
  where $\Phi$ is the time-averaged $|\Phi|$ over each track. The dashed line
  shows a power-law tail fit above a data-driven cutoff
  $\Phi_{\min}\approx 6.34\times10^{18}\,\mathrm{Mx}$, yielding exponent
  $\alpha\simeq 3.41$ (tail slope $=-(\alpha-1)\approx -2.41$ on log--log axes).
  The near-linear upper tail indicates a potential scale-free behaviour confined to the
  most flux-rich spirals.}
  \label{fig:flux_ccdf}
\end{figure}

 To assess the power-law relation, the complementary cumulative distribution function (CCDF)\citep{Resnick2007HeavyTail} \(S(x)=P(X\!\ge\!x)\) is examined. Generally, for a sample \(x_{(1)}\le\cdots\le x_{(N)}\), the empirical CCDF uses the pairs \(\bigl(x_{(i)},\,S_N(x_{(i)})=1-i/N\bigr)\). A power-law tail with density \(p(x)\propto x^{-\alpha}\) above a lower cutoff \(x_{\min}\) implies \(S(x)\propto x^{-(\alpha-1)}\). On log–log axes the points beyond \(x_{\min}\) lie approximately on a straight line with slope \(-(\alpha-1)\). The tail exponent \(\alpha\) is estimated by maximum likelihood on the subset \(x\ge x_{\min}\), with \(x_{\min}\) chosen to minimise the Kolmogorov–Smirnov distance. Following \citet{KolmogorovClausetShaliziNewman2009}, we quantify goodness of fit by the Kolmogorov–Smirnov distance
\(D=\sup_{x\ge x_{\min}}\lvert S_N(x)-S_{\mathrm{pl}}(x)\rvert\) between the empirical CCDF and the fitted power-law model. In other words a strong power-law dependence will be shown as a straight line in a CCDF log-log plot. In our case we computed the CCDF of the mean flux of every detected spiral in all its lifetime. The resulted plot is shown in Figure \ref{fig:flux_ccdf}. 

 \FloatBarrier

The CCDF of the per–spiral mean unsigned flux displays the expected curvature at moderate \(\Phi\) and a near–linear upper tail on log–log axes, indicating power–law behaviour confined to the spirals with the most magnetic flux.
Using a data–driven cutoff \(\Phi_{\min}\) chosen by minimising the Kolmogorov–Smirnov distance, the tail fit yields \(\alpha \simeq 3.4\) for \(\Phi \gtrsim 6\times10^{18}\,\mathrm{Mx}\) , the fitted line is shown against the empirical CCDF after scaling by the tail fraction to allow direct comparison. Below \(\Phi_{\min}\) the CCDF flattens 
relative to a power law, so any scale–free statement applies only to the high–flux tail. At the KS-selected cutoff \(\Phi_{\min}\simeq 6.34\times10^{18}\,\mathrm{Mx}\), the representative spiral closest to \(\Phi_{\min}\) has a fitted footprint with an equivalent circular diameter of \(\approx0.42^{\circ}\) (\(\approx5.1\,\mathrm{Mm}\)) on the surface.
With this exponent, the survival function \(S(\Phi)\equiv \Pr(\Phi'\!\ge \Phi)=1-F(\Phi)\) (i.e. the complementary CDF) scales as \(S(\Phi)\propto \Phi^{-(\alpha-1)}\approx \Phi^{-2.4}\), implying that a tenfold increase in mean flux reduces the occurrence by about \(10^{2.4}\). 
The key point is that the tail of this particular distribution is potentially compatible with a power law whereas the body is not. The near-linear CCDF tail on log–log axes points to a scale-free distribution of per spiral flux, potentially consistent with hierarchical tearing, coalescence, and secondary tearing in a plasmoid cascade the hallmark of turbulent reconnection. 

We emphasize that Figure \ref{fig:flux_ccdf} is constructed from a surface proxy: for each detected spiral we measure the unsigned magnetic flux enclosed by the fitted ellipse on the detected spiral (and time-average this quantity over the spiral lifetime).
This should not be interpreted as a direct measurement of the instantaneous plasmoid/flux-rope content in the current sheet, because the
mapping between volumetric plasmoids and surface spirals is not generally one-to-one (e.g., a single volumetric structure may map to multiple
surface imprints, and this partitioning can evolve in time, see Fig. \ref{fig:plasmoid_spiral} and \citet{Dahlin2025}).
Accordingly, the distribution in Fig. \ref{fig:flux_ccdf} is best viewed as a statistical distribution of the detected surface imprints produced in this
plasmoid-forming simulation, and provides a lower-bound, population-level proxy for comparing against theoretical expectations.
A direct quantitative comparison between volumetric plasmoid properties and their evolving surface imprints requires dedicated 3-D identification
and tracking in the current sheet and is left to a follow-up study.

\subsection{Lifetime of spirals}
\begin{figure*}[t]
  \centering
  \includegraphics[width=\textwidth,height=0.98\textheight,keepaspectratio]{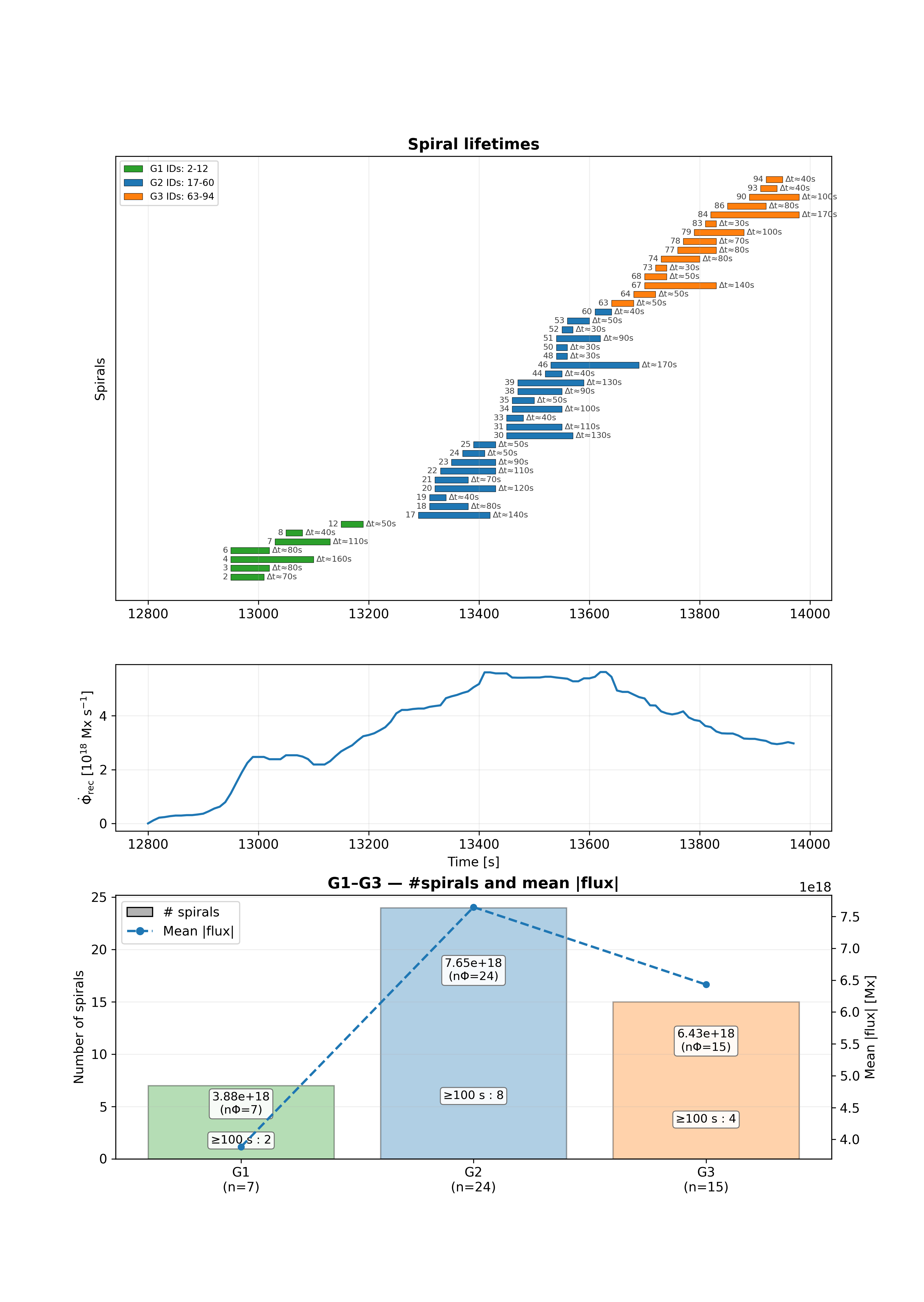}
  \caption{\textbf{Spiral lifetimes, reconnection rate, and group statistics.}
  \textit{Top:} Lifetimes of all detected spirals with duration $>20$~s, shown as horizontal bars spanning each spiral's start and end times; colors indicate groups G1--G3.
  \textit{Middle:} Time-dependent reconnection rate $\dot{\Phi}_{\rm rec}$.
  \textit{Bottom:} Per-group summary: number of spirals (bars, left axis) and mean $|\Phi|$ (dashed line, right axis). Labels report mean $|\Phi|$ (with $n_\Phi$ the number with flux estimates) and counts with lifetime $\ge100$~s.}
  \label{fig:spiral_lifetimes_reconnection_groups}
\end{figure*}

We next examine the lifetimes of the detected spirals. Figure~\ref{fig:spiral_lifetimes_reconnection_groups}(a) shows a timeline-style bar chart of spiral lifetimes, in which each horizontal bar marks a spiral’s first detection, last detection, and resulting duration. The pattern of bar appearances suggests three broad temporal intervals: an early phase G1 (IDs 1–13) with only a few short-lived spirals, a middle phase G2 (IDs 16–60) where many spirals are born in short, bursty clusters, and a late phase G3 (IDs 63–96) in which new spirals appear more sparsely  rather than in clusters. We estimate the reconnection rate by tracking time variations in the traced field-line length $L(\phi,\theta,t)$ at the lower boundary using the method introduced in \citet{Dahlin2022GuideField}. For each boundary pixel we maintain a running maximum, $L_{\max}(\phi,\theta,t)=\max_{t'\le t} L(\phi,\theta,t')$. A pixel is flagged as newly reconnected when its field line shortens \emph{permanently} by at least $40\%$ relative to its own maximum, i.e.\ when $L(\phi,\theta,t)\le 0.6\,L_{\max}(\phi,\theta,t)$ for $N_{\rm perm}$ consecutive snapshots (here $N_{\rm perm}=3$, corresponding to $30$~s for a $10$~s cadence), and each pixel is counted once. The newly reconnected flux per snapshot is then computed as $\Delta\Phi(t)=\tfrac{1}{2}\sum_{\rm new} |B_r(\phi,\theta,t)|\,{\rm d}A$, where ${\rm d}A$ is the pixel area on the photosphere and the factor $1/2$ avoids double-counting the two conjugate footpoints. Finally, the reconnection rate is obtained as $\dot{\Phi}_{\rm rec}(t)=\Delta\Phi(t)/\Delta t$, with $\Delta t$ the snapshot cadence. The time evolution of the reconnection rate, $\dot{\Phi}_{\rm rec}(t)$, is shown in the \textit{middle} panel of Fig.~\ref{fig:spiral_lifetimes_reconnection_groups}. This curve provides a simple temporal context for interpreting the spiral groupings: we associate the early-time spirals (G1) with the \emph{onset} phase, when $\dot{\Phi}_{\rm rec}$ is rising from low values, the intermediate group (G2) with the \emph{impulsive} phase, coincident with the interval of largest $\dot{\Phi}_{\rm rec}$, and the late-time spirals (G3) with the \emph{decay} phase, when $\dot{\Phi}_{\rm rec}$ decreases from its peak. The G1, G2, and G3 groups correspond to the early-, mid-, and late-phase spirals shown in Fig.~\ref{fig:mean_aspect}.
 Subsequently, G2 coincides with the interval when the straight section of the ribbons hosts the largest number of spirals, whereas G3 exhibits fewer, more sequential appearances, consistent with a waning reconnection rate toward the end of the event as the middle panel of Figure \ref{fig:spiral_lifetimes_reconnection_groups} display. 
 This phase-based grouping is used in what follows when comparing flux and occurrence statistics across the dataset. Figure~\ref{fig:spiral_lifetimes_reconnection_groups} bottom plot summarizes, by phase, (i) the number of spirals detected within the straight section of the ribbons, (ii) how many in each phase have lifetimes $>100$\,s, 
 and (iii) the phase-averaged mean unsigned flux of those spirals. In the onset phase (G1) the spiral count is comparatively small and the mean flux is the lowest, consistent with a gradual build-up preceding vigorous reconnection.  
 The impulsive phase (G2) shows the largest number of detected spirals, the highest mean flux, and the greatest count of long-lived ($>100$\,s) spirals, reflecting peak energy release. Finally, during the decay phase (G3) both the spiral count and mean flux decrease, as does the number of long-lived spirals, consistent with a decaying reconnection rate. We return to the physical interpretation of these statistical trends in Section~\ref{sec:conclusions}.

\section{Discussion}
\label{sec:discussion}
The high–Lundquist-number 3-D simulation of \citet{Joel_2022aApJ...932...94D} produces a highly fragmented current sheet and numerous plasmoids, which motivated us to develop an automated method to study their imprint on the flare ribbons in detail.

We used the CDM method introduced by \citet{Mason2022ApJ...937L..19M} for coronal hole boundaries, but applied it slightly differently by breaking up the analysis into multiple passes at different scales.  
By performing a three-pass CDM over nested radius bands, we capture the structured
character of flare-ribbon boundaries across multiple spatial scales and thereby detect spiral structures consistently over those scales. Subsequently, we cluster the retained points (those with elevated correlation dimension, \(D>1\)) using \textsc{DBSCAN} to isolate coherent spiral segments, and we fit a MAEE to each cluster. This yielded spiral detections consistently across the three spatial scales probed by the CDM passes and helped offset the natural tendency for small-scale features to be found within larger-scale ones. The method systematically detected spiral-like segments along the uniformly sampled boundaries. 
We find that all detected spirals follow a similar temporal trend in the aspect ratio of their fitted ellipses. Initially, each spiral starts out flat to the ribbon, characterised by a high aspect ratio ellipse, before growing in width normal to the ribbon front, so that their ellipse becomes more circular. This systematic evolution was shown in \citet{Dahlin2025} (see e.g. their Fig. 14) to correspond to the initial growth of the plasmoid within the layer, which evolves the spiral from high to lower aspect ratio. Towards the end of their lifetimes, their ellipse then becomes more elongated once more, with the major axis now more aligned normal to the ribbon front. This corresponds to the plasmoid exiting the current layer and detaching from the flux surface that defines the ribbon front \citep{Wyper_Pontin2021ApJ...920..102W,Dahlin2025}. Confirming this evolution also gives us confidence that the features identified by our method are indeed the result of plasmoid dynamics in the current layer. We also noted a general trend for the ellipses to be more circular on average throughout their lifetimes in the later stages of the flare in this simulation.

\par To further follow the kinematics of the ribbons, we also tracked the time-dependent separation of the two flare ribbons by fitting the average 
ribbon curve at each snapshot. The mean vertical speed at the locations of maximum separation is \(\sim 40\kms\), followed by a gradual decline as the event enters its decay phase. These values are comparable to ribbon–separation speeds reported in observations (e.g., \citealp{Asai2004ApJ611557,Hinterreiter2018SoPh29386,Chen2012ChinSciBull57}). Using the fitted ribbon curve as a moving reference, we find that the spirals detected along straight sections are \emph{kinematically locked} to the ribbon: their across–ribbon (normal) offsets remain small compared with their along–ribbon drift, and their centroids co-migrate with the ribbon as it advances. In practice, none of the spirals overtakes the ribbon front or detaches from  it; even the early examples, which show slightly larger motion, remain embedded in the ribbon. This behaviour is natural as
the spirals mark the photospheric footpoints of plasmoids forming in the flare current sheet so should be expected to broadly move with it \citep{Wyper_Pontin2021ApJ...920..102W,Dahlin2025}. 

Along the straight sections of the ribbons, the spirals do however exhibit a clear, coherent \emph{lateral} (along–ribbon) drift: all spirals within a given ribbon move in the same sense, away from their respective hooks. This coherent drift is consistent with the drift postulated in \citet{Wyper_Pontin2021ApJ...920..102W} and demonstrated dynamically in \citet{Dahlin2025}, arising from field lines leaving given plasmoids slipping within the current layer as the plasmoids propagate and are ejected, in this case downwards towards the flare loops. Superposed on this  trend is substantial diversity in the magnitude and time profile of the lateral drift speeds, which fall in the range 10-800 km\,s$^{-1}$, and   
evolve differently from spiral to spiral. Such variability is plausibly linked to changes in the reconnection environment (e.g. the guide field, the height at which plasmoids form, and the length of the reconnected field lines to which they map), each of which  would be expected to modulate the apparent footpoint motion along the ribbon. Importantly, this drift range is substantially below the surface Alfv\'{e}n speed ($v_a \approx 3000$ km\,$^{-1}$, with the range corresponding to $0.003-0.27 v_a$). This highlights that the collective drift of the plasmoid foot points, i.e. the spirals, is likely different to the general connectivity change associated with the average field line slippage occurring globally due to 3D reconnection within the current layer, which tends to be close to or even above $v_a$ \citep[e.g.][]{Lorincik2025NatAs...9...45L}. Of course, the latter must be occurring at the same time as it supports the general motion of the flare ribbons, but the key point is that plasmoids need not necessarily follow the same motion as field lines anchored on the surface (against which slipping reconnection is defined), since the plasmoids themselves are embedded within and slipping through the non-ideal flare current sheet region. In future work we will further explore this duality of connection change. 
Taken together, these results indicate that tracking fine–structure kinematics at the surface has the potential to provide a window onto the three–dimensional behaviour of plasmoids in the flare current sheet. In particular, systematic trends in the along–ribbon drift (direction, acceleration, and characteristic speeds) offer a potential diagnostic to distinguish different plasmoid populations (e.g.\ upward vs.\ downward moving), while aspect ratio evolution and magnetic flux content were shown to vary across the lifetime of the flare and as such could potentially be used to infer more broad current sheet properties such as the reconnection guide field and flare energy release. In this work we aimed to establish the methodology for identifying these systematic trends. In future work these results will be combined with the volumetric plasmoid properties to address whether a quantitative mapping between surface trends and current sheet dynamics can be established.

\par As first step in this direction, however, we investigated the statistics of per-spiral magnetic flux (the surface flux within their fitted ellipse, \(\Phi\)), which provides a lower bound for the axial flux threading the associated plasmoid in the current layer, Fig. \ref{fig:flux_ccdf}. A CCDF was constructed to look for an power law distribution in the occurrence rate vs magnetic flux. We find evidence for a potential power law, but only in a tail above a threshold value, \(\Phi_{\min}\).
Fitting just that tail yields \(p(\Phi)\propto \Phi^{-\alpha}\) with
\(\alpha\simeq 3.4\) (CCDF slope $\approx-2.41$), but this estimate is conditional on the chosen tail
interval and is sensitive to \(\Phi_{\min}\) and sample size. We therefore interpret the result as evidence for a \emph{potentially} scale-free tail component, not as a universal scaling across the whole distribution. Placing the power law tail in context, most theoretical power laws are derived for
\emph{2-D} plasmoids and are written in terms of the island’s reconnected
magnetic flux $\psi$: stochastic–chain and kinetic models predict
$f(\psi)\!\propto\!\psi^{-2}$ or, over an intermediate range,
$f(\psi)\!\propto\!\psi^{-1}$ \citep{Uzdensky2010PRL,
PoP2012_LoureiroUzdenskyConfirm,HuangBhattacharjee2012PRL}.
For \emph{3-D} flux ropes, a maximum–entropy treatment predicts a flux distribution
$f(\psi)\!\propto\!\psi^{-3}$ with an exponential cutoff at large $\psi$
\citep{LingamComisso2018PoP}. 
Our measured tail is therefore steeper than
the classic 2-D flux–distribution predictions and close to the 3-D
$\psi^{-3}$ scaling. 
That being said, our measured flux is a time–averaged imprint of the guide field through the plasmoids within the flare current layer not the instantaneous in–sheet $\psi$. Therefore, the distribution shown in Fig. \ref{fig:flux_ccdf} should be interpreted as a statistical distribution of detected surface imprints (a lower-bound proxy), rather than as a direct measurement of an in-sheet plasmoid-flux distribution. Despite this, our results are still suggestive that the plasmoids within the flare current layer follow a power-law distribution in some form. Exploring the plasmoid distribution within the volume, as well as its dependence upon resolution, will be the topic of future work.

\par Lastly, in  Sect. 5.4 we showed that the appearance of the spirals in the straight section could be roughly split in time based on their evolution characteristics into three phases: 
an onset phase (G1), an impulsive phase (G2), and a decay phase (G3).  
During the onset (G1) only a few, short–lived spirals appear and their mean $|\Phi|$ is the lowest. In the impulsive phase (G2) both the number of spirals and the fraction of long–lived spirals increase sharply, and the phase–averaged mean $|\Phi|$ reaches a maximum. In the decay (G3) the counts fall again and the mean $|\Phi|$ decreases relative to G2. Overall, the lifetime distribution and the occurrence rate both peak during G2, a result worth highlighting for observational studies.
In \citet{Joel_2022aApJ...932...94D}, and more specifically in  Fig.~7, the guide–field proxy $-B_\phi/|B_r|$ decreases monotonically while the reconnection rate (red curve, in $10^{18}$\,Mx\,s$^{-1}$) rises to a peak near $t\!\sim\!13.0$~ks and then declines. When the guide field is strongest (early times), the current sheet produces few, weak spirals (G1). As the guide field weakens and the reconnection rate surges, the sheet fragments more vigorously, feeding flux into many plasmoids simultaneously. This coincides with the G2 maximum in counts, lifetimes, and mean $|\Phi|$. Once the reconnection rate subsides (G3), fewer new spirals are identified and the mean per–spiral flux drops. This does not necessarily represent a drop in the number of plasmoids within the current layer (which continues to lengthen during this phase), rather it reflects the fact that the guide field through the main flare current layer has substantially reduced by this phase, and consequently so too has the guide field threading the plasmoids, appearing as a drop in the per-spiral flux and a general reduction in the numbers of spirals identified. Together, these results highlight that (in the straight sections of the ribbons) spirals are expected to be most prevalent in impulsive phase of the flare, when the reconnection rate is intense, but a non-negligible guide field still persists within the current layer.

\section{Conclusions}
\label{sec:conclusions}
In summary, we have developed a method for identifying and tracking spiral and wave-like sub-structures in flare ribbons derived from the simulation of \citet{Dahlin2025} which are associated with plasmoids within the flare current layer. The qualitative and quantitative statistical trends show differences throughout the lifetime of the simulated flare, hinting that such ribbon sub-structure provides a practical surface diagnostic of the reconnection process and its temporal phases. The method presented here provides an automated way to detect and track the geometrically complex segments of ribbon-front (or analogous)
boundary contours using a purely geometric criterion (local correlation-dimension mapping) followed by clustering and geometric fitting.
A key strength is that the approach is mechanism-agnostic: it does not assume a unique physical process responsible for the fine structure,
but rather quantifies boundary complexity that can arise under a range of flare conditions. In the present paper we validate the pipeline in a
simulation setting where spiral imprints on the lower boundary are demonstrably linked to plasmoids in the 3D current sheet.

The method also has limitations that should be kept in mind when interpreting the results presented here.
First, performance is tied to the quality and consistency of the boundary extraction: if the ribbon front cannot be segmented consistently
across time, the resulting contours can fragment, which directly impacts the CDM estimates and the subsequent clustering. In the present
simulation, however, this step is comparatively well controlled because the contours are extracted from field-line-length maps in a clean
and physically interpretable setting. Second,
CDM relies on an approximately linear scaling of \(\log C(r)\) versus \(\log r\) over a finite range of radii; therefore the radius bands
must be chosen to lie above the effective resolution limit and within the scale range where the boundary morphology is meaningfully sampled.
A practical challenge in applying the method is therefore to choose scale ranges that isolate the spiral/wave-like substructure without
instead responding to larger-scale ribbon shape or to very small-scale contour irregularities. Finally, although we show robustness to
parameter choices within the ranges explored here, the clustering stage (DBSCAN in this work) still requires user-defined parameters, and
different datasets will require effort to define parameter values that yield accurate results. In this sense, the results presented here are
trustworthy within the tested parameter ranges and for this particular dataset, but they remain subject to the caveats associated with
contour extraction, scale selection, and clustering choices.

More broadly, the method detects complex projected boundary structure and does not, by itself, provide a unique physical attribution:
different 3D processes (e.g. the tearing or Kelvin--Helmholtz instabilities) can in theory lead to boundary perturbations with similar
projected complexity, and morphology alone is insufficient to discriminate between candidate trigger mechanisms without additional diagnostics.
What is clear, however, is that regardless of the initial trigger, the presence of plasmoids in the current layer leads to spiral-like
features in the ribbon front, which our method has been able to identify and characterise for this particular dataset. We therefore view
this study as a proof of concept that establishes an end-to-end, reproducible pipeline and a quantitative catalogue of detected complex
boundary segments. 

Applying the method to observational data introduces additional challenges beyond those considered in this paper. In particular, the
reliable extraction of ribbon-front boundaries from intensity images is more difficult than in the present simulation-derived field-line-length
maps, because observational fronts can evolve against structured backgrounds, may be less sharply defined in space and time, and can be
artificially fragmented by the boundary-extraction procedure itself. A recent observational application by \citet{CorchadoAlbelo2026}
demonstrates that such analyses are feasible in IRIS data, but also highlights that the robustness of the inferred ribbon complexity depends
sensitively on the contrast of the flare-ribbon bright leading edge, the spatial resolution and cadence of the observations, and the stability of the extraction method against noise and saturation effects. The method is therefore expected to be most reliable when the ribbon front is
sufficiently well resolved and can be tracked consistently from frame to frame. 

In future work we aim to better quantify the link between 3D plasmoids and their spiral-like projections in the ribbons by considering the volumetric evolution of the plasmoids, as well as the comparative statistical properties of the current layer.
While our implementation uses $L$-maps, the workflow can be adapted to other connectivity proxies available from observations of real flare ribbons, enabling direct application to flare ribbon data. This is also the subject of ongoing work.



\begin{acknowledgments}
We thank the referees for valuable comments leading to the improvement of this paper, and Alex Russell, Emily Mason and Vadim Uritsky for illuminating and helpful discussions. This work was supported by Leverhulme Trust project grant RPG-2023-288. Lyndsay Fletcher is grateful for support from UK Research and Innovation’s Science and Technology Facilities Council,
under grant award number ST/X000990/1.
\end{acknowledgments}

\appendix
\setcounter{equation}{0}
\renewcommand{\theequation}{A\arabic{equation}}
\renewcommand{\theHequation}{A.\arabic{equation}}
\section{Comparison of DBSCAN and HDBSCAN for Spiral Detection}
\label{app:dbscan_hdbscan}

Because the clustering stage of our pipeline is applied only after the CDM has already selected candidate high-complexity core points, the role of the clustering algorithm is not to discover arbitrary structure in the full ribbon-front contour, but rather to group these pre-selected core points into coherent detections. In the simulation analysed here, these candidate detections are typically compact and spatially isolated along the ribbon front, separated by extended smoother sections with few or no CDM-selected points. This is precisely the regime in which DBSCAN is expected to perform well: it groups nearby core points into distinct clusters while naturally rejecting sparse outliers as noise.

To verify that our results are not sensitive to the choice of density-based clustering algorithm, we repeated the full clustering stage using HDBSCAN in place of DBSCAN. The comparison was designed so that the two methods were treated as consistently as possible. In particular, the upstream steps of the pipeline were kept identical in all cases: the same field-line-length maps were used, the same contour extraction was applied, the same three CDM radius bands were used, and the same local dimension value threshold for selecting core points was imposed. Likewise, the downstream post-processing was kept unchanged. Thus, the only step that differs between the two cases is the clustering backend itself.

For DBSCAN we used the same parameters as in the main analysis, namely \(\epsilon = 0.06\) and \texttt{min\_samples} \(=5\). For HDBSCAN, we adopted a directly comparable but not specially tuned setup: \texttt{min\_samples} was also set to \(5\), while \texttt{min\_cluster\_size} was set to \(\max(2\times \texttt{min\_samples},\,8)=10\). We kept the HDBSCAN cluster-selection epsilon at its default value (\(0\)), rather than adjusting it to mimic DBSCAN. In this way, the HDBSCAN comparison acts as a genuine test of whether a hierarchical density-based variant changes the recovered detections for this dataset.
\begin{figure*}
    \centering
    \includegraphics[width=\textwidth]{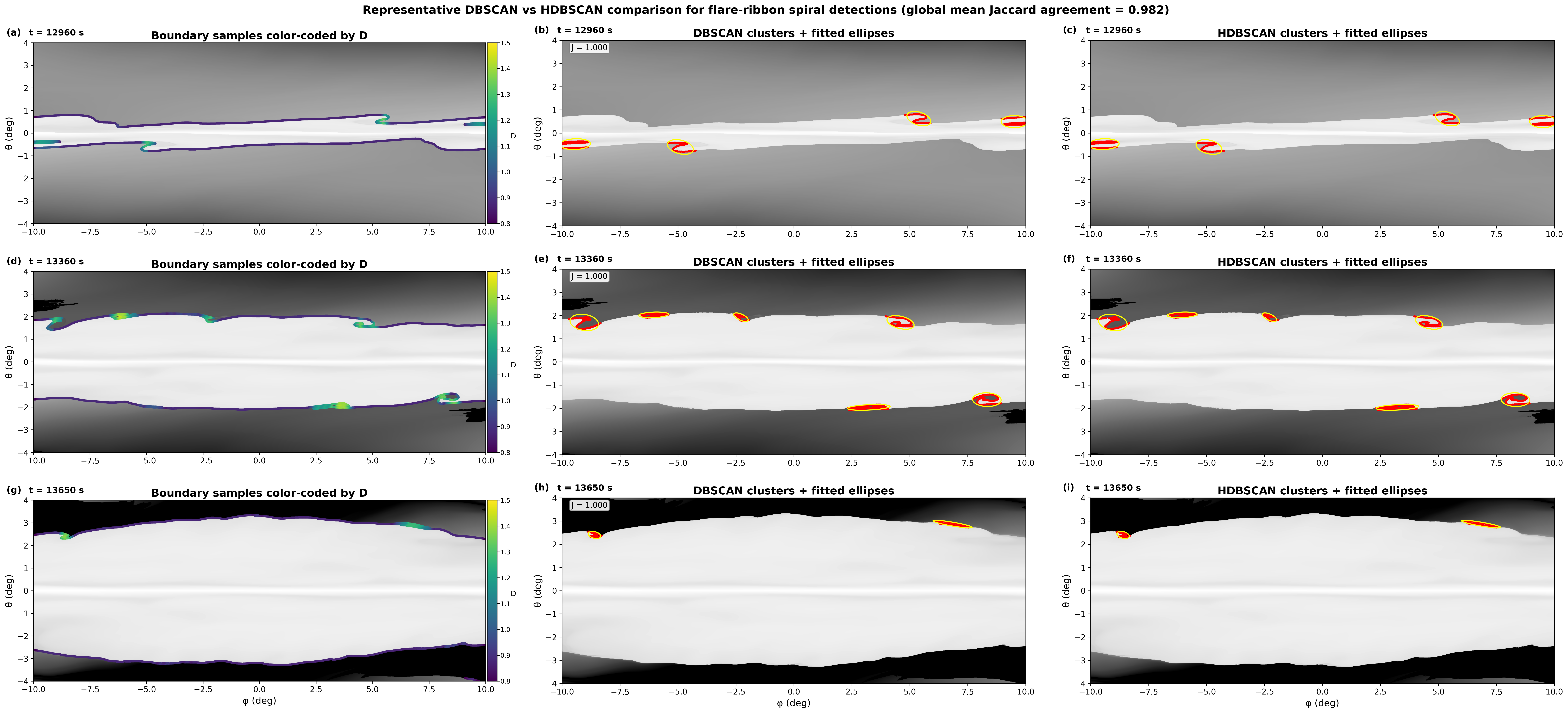}
    \caption{Representative comparison of spiral detections obtained with DBSCAN and HDBSCAN. Each row shows a different simulation snapshot (\(t=12960\), \(13360\), and \(13650\)~s). In each row, the left panel displays the ribbon-front boundary samples colored by the local correlation-dimension estimate \(D\), the middle panel shows the final detections obtained with DBSCAN, and the right panel shows the corresponding detections obtained with HDBSCAN.}
    \label{fig:appendix_dbscan_hdbscan}
\end{figure*}
Figure~\ref{fig:appendix_dbscan_hdbscan} shows representative snapshots from the  simulation. In each row, the left panel shows the ribbon-front boundary samples coloured by the local correlation-dimension \(D\), the middle panel shows the final DBSCAN detections, and the right panel shows the corresponding HDBSCAN detections. The qualitative agreement is very strong: in all representative cases the same spiral-like structures are isolated and fitted by nearly identical ellipses.

To quantify this agreement more systematically, we compared the final detections frame by frame after the full post-processing stage. Detections from DBSCAN and HDBSCAN were matched. From these matches we computed a Jaccard agreement,
\begin{equation}
J = \frac{N_{\mathrm{match}}}{N_{\mathrm{DBSCAN}} + N_{\mathrm{HDBSCAN}} - N_{\mathrm{match}}},
\end{equation}
where \(N_{\mathrm{match}}\) is the number of matched detections in a frame, and \(N_{\mathrm{DBSCAN}}\) and \(N_{\mathrm{HDBSCAN}}\) are the total numbers of detections returned by the two methods. Averaged over the tested frames, we obtain a mean Jaccard agreement of \(J=0.982\), showing that the two clustering algorithms yield essentially identical results for the present dataset.

We therefore conclude that, for the flare-ribbon boundary structures studied here, DBSCAN is a sufficient and well-justified choice. Its simplicity and transparency make it an appropriate baseline for this proof-of-concept study, while the HDBSCAN comparison demonstrates that our results are not materially altered by adopting a more sophisticated hierarchical density-based variant. This does not imply that DBSCAN will always be optimal for every dataset; in cases where detections are much less well separated or where cluster densities vary strongly, hierarchical approaches such as HDBSCAN may prove advantageous.
\bibliographystyle{aasjournal}
\bibliography{bibliography}

\begin{thebibliography}{}
\expandafter\ifx\csname natexlab\endcsname\relax\def\natexlab#1{#1}\fi
\providecommand{\url}[1]{\href{#1}{#1}}
\providecommand{\dodoi}[1]{doi:~\href{http://doi.org/#1}{\nolinkurl{#1}}}
\providecommand{\doeprint}[1]{\href{http://ascl.net/#1}{\nolinkurl{http://ascl.net/#1}}}
\providecommand{\doarXiv}[1]{\href{https://arxiv.org/abs/#1}{\nolinkurl{https://arxiv.org/abs/#1}}}

\bibitem[{Asai {et~al.}(2004)Asai, Yokoyama, Shimojo, Masuda, Kurokawa, \&
  Shibata}]{Asai2004ApJ611557}
Asai, A., Yokoyama, T., Shimojo, M., {et~al.} 2004, The Astrophysical Journal,
  611, 557, \dodoi{10.1086/422159}

\bibitem[{{Aulanier} {et~al.}(2012){Aulanier}, {Janvier}, \&
  {Schmieder}}]{Aulanier2012A&A...543A.110A}
{Aulanier}, G., {Janvier}, M., \& {Schmieder}, B. 2012, \aap, 543, A110,
  \dodoi{10.1051/0004-6361/201219311}

\bibitem[{Battaglia {et~al.}(2009)Battaglia, Fletcher, \&
  Benz}]{Battaglia2009AA}
Battaglia, M., Fletcher, L., \& Benz, A.~O. 2009, Astronomy \& Astrophysics,
  498, 891, \dodoi{10.1051/0004-6361/200811196}

\bibitem[{{Beg} {et~al.}(2022){Beg}, {Russell}, \& {Hornig}}]{Beg2022}
{Beg}, R., {Russell}, A. J.~B., \& {Hornig}, G. 2022, \apj, 940, 94,
  \dodoi{10.3847/1538-4357/ac8eb6}

\bibitem[{Benz(2017)}]{Benz2017LRSP}
Benz, A.~O. 2017, Living Reviews in Solar Physics, 14, 2,
  \dodoi{10.1007/s41116-017-0002-1}

\bibitem[{Canny(1986)}]{Canny1986}
Canny, J. 1986, IEEE Transactions on Pattern Analysis and Machine Intelligence,
  PAMI-8, 679, \dodoi{10.1109/TPAMI.1986.4767851}

\bibitem[{{Carmichael}(1964)}]{Carmichael_1964}
{Carmichael}, H. 1964, NASA Special Publication, 50, 451

\bibitem[{{Che} \& {Zank}(2020)}]{Che2020}
{Che}, H., \& {Zank}, G.~P. 2020, \apj, 889, 11,
  \dodoi{10.3847/1538-4357/ab5d3b}

\bibitem[{Chen {et~al.}(2012)Chen, Deng, Fang, \& Liu}]{Chen2012ChinSciBull57}
Chen, P.~F., Deng, Y.~Y., Fang, C., \& Liu, Y. 2012, Chinese Science Bulletin,
  57, 1392, \dodoi{10.1007/s11434-011-4829-9}

\bibitem[{Clauset {et~al.}(2009)Clauset, Shalizi, \&
  Newman}]{KolmogorovClausetShaliziNewman2009}
Clauset, A., Shalizi, C.~R., \& Newman, M. E.~J. 2009, SIAM Review, 51, 661,
  \dodoi{10.1137/070710111}

\bibitem[{Corchado~Albelo {et~al.}(2026)Corchado~Albelo, Kazachenko, French,
  Uritsky, Mason, Tamburri, Yadav, \& Lynch}]{CorchadoAlbelo2026}
Corchado~Albelo, M.~F., Kazachenko, M.~D., French, R.~J., {et~al.} 2026, arXiv
  e-prints, arXiv:2602.20470.
\newblock \doarXiv{2602.20470}

\bibitem[{{Corchado Albelo} {et~al.}(2024){Corchado Albelo}, {Kazachenko}, \&
  {Lynch}}]{CorchadoAlbelo2024}
{Corchado Albelo}, M.~F., {Kazachenko}, M.~D., \& {Lynch}, B.~J. 2024, \apj,
  965, 16, \dodoi{10.3847/1538-4357/ad25f4}

\bibitem[{{Dahlin} {et~al.}(2022){Dahlin}, {Antiochos}, {Qiu}, \&
  {DeVore}}]{Joel_2022aApJ...932...94D}
{Dahlin}, J.~T., {Antiochos}, S.~K., {Qiu}, J., \& {DeVore}, C.~R. 2022, \apj,
  932, 94, \dodoi{10.3847/1538-4357/ac6e3d}

\bibitem[{Dahlin {et~al.}(2022)Dahlin, Antiochos, Qiu, \&
  DeVore}]{Dahlin2022GuideField}
Dahlin, J.~T., Antiochos, S.~K., Qiu, J., \& DeVore, C.~R. 2022, The
  Astrophysical Journal, 932, 94, \dodoi{10.3847/1538-4357/ac6e3d}

\bibitem[{{Dahlin} {et~al.}(2025){Dahlin}, {Antiochos}, {Wyper}, {Qiu}, \&
  {DeVore}}]{Dahlin2025}
{Dahlin}, J.~T., {Antiochos}, S.~K., {Wyper}, P.~F., {Qiu}, J., \& {DeVore},
  C.~R. 2025, \apj, 993, 31, \dodoi{10.3847/1538-4357/ae03c5}

\bibitem[{{Dahlin} {et~al.}(2022){Dahlin}, {DeVore}, \&
  {Antiochos}}]{Stitch2022ApJ...941...79D}
{Dahlin}, J.~T., {DeVore}, C.~R., \& {Antiochos}, S.~K. 2022, \apj, 941, 79,
  \dodoi{10.3847/1538-4357/ac9e5a}

\bibitem[{{Dud{\'\i}k} {et~al.}(2025){Dud{\'\i}k}, {Aulanier},
  {L{\"o}rin{\v{c}}{\'\i}k}, \& {Zemanov{\'a}}}]{Dudik2025}
{Dud{\'\i}k}, J., {Aulanier}, G., {L{\"o}rin{\v{c}}{\'\i}k}, J., \&
  {Zemanov{\'a}}, A. 2025, \solphys, 300, 139,
  \dodoi{10.1007/s11207-025-02549-2}

\bibitem[{Dud{\'\i}k {et~al.}(2014)Dud{\'\i}k, Janvier, Aulanier, Del~Zanna,
  Karlick{\'y}, Mason, \& Schmieder}]{Dudik2014ApJ}
Dud{\'\i}k, J., Janvier, M., Aulanier, G., {et~al.} 2014, The Astrophysical
  Journal, 784, 144, \dodoi{10.1088/0004-637X/784/2/144}

\bibitem[{Dud{\'\i}k {et~al.}(2016)Dud{\'\i}k, Polito, Janvier, Mulay,
  Karlick{\'y}, Aulanier, Del~Zanna, Dzif{\v c}{\'a}kov{\'a}, Mason, \&
  Schmieder}]{Dudik2016ApJ}
Dud{\'\i}k, J., Polito, V., Janvier, M., {et~al.} 2016, The Astrophysical
  Journal, 823, 41, \dodoi{10.3847/0004-637X/823/1/41}

\bibitem[{Ester {et~al.}(1996)Ester, Kriegel, Sander, \&
  Xu}]{DBSCANEsterKriegelSanderXu1996}
Ester, M., Kriegel, H., Sander, J., \& Xu, X. 1996, in Proceedings of the
  Second International Conference on Knowledge Discovery and Data Mining
  (KDD-96) (Portland, OR, USA: AAAI Press), 226--231

\bibitem[{Fletcher \& Hudson(2008)}]{FletcherHudson2008ApJ}
Fletcher, L., \& Hudson, H.~S. 2008, The Astrophysical Journal, 675, 1645,
  \dodoi{10.1086/527044}

\bibitem[{Fletcher {et~al.}(2011)Fletcher, Dennis, Hudson, Krucker, Phillips,
  Veronig, Battaglia, Bone, Caspi, Chen, Gallagher, Grigis, Ji, Liu, Milligan,
  \& Temmer}]{Fletcher2011SSRv}
Fletcher, L., Dennis, B.~R., Hudson, H.~S., {et~al.} 2011, Space Science
  Reviews, 159, 19, \dodoi{10.1007/s11214-010-9701-8}

\bibitem[{Grassberger \& Procaccia(1983)}]{GrassbergerProcaccia1983PhysicaD}
Grassberger, P., \& Procaccia, I. 1983, Physica D: Nonlinear Phenomena, 9, 189,
  \dodoi{10.1016/0167-2789(83)90298-1}

\bibitem[{Guidoni \& Longcope(2011)}]{Guidoni2011DensityEnhancements}
Guidoni, S.~E., \& Longcope, D.~W. 2011, The Astrophysical Journal, 730, 90,
  \dodoi{10.1088/0004-637X/730/2/90}

\bibitem[{Hinterreiter {et~al.}(2018)Hinterreiter, Veronig, Thalmann,
  Tschernitz, \& P{\"o}tzi}]{Hinterreiter2018SoPh29386}
Hinterreiter, J., Veronig, A.~M., Thalmann, J.~K., Tschernitz, J., \&
  P{\"o}tzi, W. 2018, Solar Physics, 293, 86, \dodoi{10.1007/s11207-018-1253-1}

\bibitem[{{Hirayama}(1974)}]{Hirayama_1974}
{Hirayama}, T. 1974, \solphys, 34, 323, \dodoi{10.1007/BF00153671}

\bibitem[{Holman {et~al.}(2011)Holman, Aschwanden, Aurass, Battaglia, Grigis,
  Kontar, Liu, Saint-Hilaire, \& Zharkova}]{Holman2011SSR}
Holman, G.~D., Aschwanden, M.~J., Aurass, H., {et~al.} 2011, Space Science
  Reviews, 159, 107, \dodoi{10.1007/s11214-010-9680-9}

\bibitem[{Huang \& Bhattacharjee(2012)}]{HuangBhattacharjee2012PRL}
Huang, Y.-M., \& Bhattacharjee, A. 2012, Phys. Rev. Lett., 109, 265002,
  \dodoi{10.1103/PhysRevLett.109.265002}

\bibitem[{{Huang} \& {Bhattacharjee}(2016)}]{Huang2016}
{Huang}, Y.-M., \& {Bhattacharjee}, A. 2016, \apj, 818, 20,
  \dodoi{10.3847/0004-637X/818/1/20}

\bibitem[{{Janvier} {et~al.}(2013){Janvier}, {Aulanier}, {Pariat}, \&
  {D{\'e}moulin}}]{Janvier2013A&A...555A..77J}
{Janvier}, M., {Aulanier}, G., {Pariat}, E., \& {D{\'e}moulin}, P. 2013, \aap,
  555, A77, \dodoi{10.1051/0004-6361/201321164}

\bibitem[{{Jing} {et~al.}(2016){Jing}, {Xu}, {Cao}, {Liu}, {Gary}, \&
  {Wang}}]{Jing2016}
{Jing}, J., {Xu}, Y., {Cao}, W., {et~al.} 2016, Scientific Reports, 6, 24319,
  \dodoi{10.1038/srep24319}

\bibitem[{{Karpen} {et~al.}(2012){Karpen}, {Antiochos}, \&
  {DeVore}}]{Karpen_etal2012}
{Karpen}, J.~T., {Antiochos}, S.~K., \& {DeVore}, C.~R. 2012, \apj, 760, 81,
  \dodoi{10.1088/0004-637X/760/1/81}

\bibitem[{{Karpen} {et~al.}(2024){Karpen}, {Kumar}, {Wyper}, {DeVore}, \&
  {Antiochos}}]{Karpen2024}
{Karpen}, J.~T., {Kumar}, P., {Wyper}, P.~F., {DeVore}, C.~R., \& {Antiochos},
  S.~K. 2024, \apj, 966, 27, \dodoi{10.3847/1538-4357/ad2eaa}

\bibitem[{{Kazachenko} {et~al.}(2022){Kazachenko}, {Albelo-Corchado},
  {Tamburri}, \& {Welsch}}]{Kazachenko2022}
{Kazachenko}, M.~D., {Albelo-Corchado}, M.~F., {Tamburri}, C.~A., \& {Welsch},
  B.~T. 2022, \solphys, 297, 59, \dodoi{10.1007/s11207-022-01987-6}

\bibitem[{Kazachenko {et~al.}(2017)Kazachenko, Lynch, Welsch, \&
  Sun}]{Kazachenko2017ApJ_RibbonDB}
Kazachenko, M.~D., Lynch, B.~J., Welsch, B.~T., \& Sun, X. 2017, The
  Astrophysical Journal, 845, 49, \dodoi{10.3847/1538-4357/aa7ed6}

\bibitem[{{Kopp} \& {Pneuman}(1976)}]{Kopp_etal1976}
{Kopp}, R.~A., \& {Pneuman}, G.~W. 1976, \solphys, 50, 85,
  \dodoi{10.1007/BF00206193}

\bibitem[{{Kumar} {et~al.}(2021){Kumar}, {Karpen}, {Antiochos}, {Wyper},
  {DeVore}, \& {Lynch}}]{Kumar2021}
{Kumar}, P., {Karpen}, J.~T., {Antiochos}, S.~K., {et~al.} 2021, \apj, 907, 41,
  \dodoi{10.3847/1538-4357/abca8b}

\bibitem[{Li \& Zhang(2015)}]{LiZhang2015ApJL}
Li, T., \& Zhang, J. 2015, The Astrophysical Journal Letters, 804, L8,
  \dodoi{10.1088/2041-8205/804/1/L8}

\bibitem[{Lingam \& Comisso(2018)}]{LingamComisso2018PoP}
Lingam, M., \& Comisso, L. 2018, Physics of Plasmas, 25, 012114,
  \dodoi{10.1063/1.5020887}

\bibitem[{Liu {et~al.}(2013)Liu, Chen, \& Petrosian}]{Liu2013PlasmoidEjections}
Liu, W., Chen, Q., \& Petrosian, V. 2013, The Astrophysical Journal, 767, 168,
  \dodoi{10.1088/0004-637X/767/2/168}

\bibitem[{L{\"o}rin{\v c}{\'\i}k {et~al.}(2019)L{\"o}rin{\v c}{\'\i}k,
  Dud{\'\i}k, Aulanier, Zemanov{\'a}, D{\'e}moulin, Schmieder, \&
  Del~Zanna}]{Lorincik2019ApJ}
L{\"o}rin{\v c}{\'\i}k, J., Dud{\'\i}k, J., Aulanier, G., {et~al.} 2019, The
  Astrophysical Journal, 881, 68, \dodoi{10.3847/1538-4357/ab298f}

\bibitem[{{L{\"o}rin{\v{c}}{\'\i}k}
  {et~al.}(2025{\natexlab{a}}){L{\"o}rin{\v{c}}{\'\i}k}, {Dud{\'\i}k}, {Sainz
  Dalda}, {Aulanier}, {Polito}, \& {De Pontieu}}]{Lorinckik2025}
{L{\"o}rin{\v{c}}{\'\i}k}, J., {Dud{\'\i}k}, J., {Sainz Dalda}, A., {et~al.}
  2025{\natexlab{a}}, Nature Astronomy, 9, 45,
  \dodoi{10.1038/s41550-024-02396-4}

\bibitem[{{L{\"o}rin{\v{c}}{\'\i}k}
  {et~al.}(2025{\natexlab{b}}){L{\"o}rin{\v{c}}{\'\i}k}, {Dud{\'\i}k}, {Sainz
  Dalda}, {Aulanier}, {Polito}, \& {De Pontieu}}]{Lorincik2025NatAs...9...45L}
---. 2025{\natexlab{b}}, Nature Astronomy, 9, 45,
  \dodoi{10.1038/s41550-024-02396-4}

\bibitem[{{Loureiro} {et~al.}(2007){Loureiro}, {Schekochihin}, \&
  {Cowley}}]{Loureiro2007}
{Loureiro}, N.~F., {Schekochihin}, A.~A., \& {Cowley}, S.~C. 2007, \phpl, 14,
  100703, \dodoi{10.1063/1.2783986}

\bibitem[{Loureiro \& Uzdensky(2012)}]{PoP2012_LoureiroUzdenskyConfirm}
Loureiro, N.~F., \& Uzdensky, D.~A. 2012, Phys. Plasmas, 19, 042303,
  \dodoi{10.1063/1.3703318}

\bibitem[{{Mason} \& {Uritsky}(2022)}]{Mason2022ApJ...937L..19M}
{Mason}, E.~I., \& {Uritsky}, V.~M. 2022, \apjl, 937, L19,
  \dodoi{10.3847/2041-8213/ac9124}

\bibitem[{McKenzie \& Hudson(1999)}]{McKenzieHudson1999_ApJL_SADs}
McKenzie, D.~E., \& Hudson, H.~S. 1999, The Astrophysical Journal Letters, 519,
  L93, \dodoi{10.1086/312110}

\bibitem[{Miklenic {et~al.}(2007)Miklenic, Veronig, Vr{\v{s}}nak, \&
  Hanslmeier}]{Miklenic2007_AA}
Miklenic, C.~H., Veronig, A.~M., Vr{\v{s}}nak, B., \& Hanslmeier, A. 2007,
  Astronomy \& Astrophysics, 461, 697, \dodoi{10.1051/0004-6361:20065751}

\bibitem[{Milligan(2011)}]{Milligan2011_ApJ_NonthermalBroadening}
Milligan, R.~O. 2011, The Astrophysical Journal, 740, 70,
  \dodoi{10.1088/0004-637X/740/2/70}

\bibitem[{Patel {et~al.}(2020)Patel, Pant, Srivastava, Banerjee, Stangalini, \&
  Uddin}]{Patel2020_AA_PostCMEPlasmoids}
Patel, R., Pant, V., Srivastava, A.~K., {et~al.} 2020, Astronomy \&
  Astrophysics, \dodoi{10.1051/0004-6361/202039000}

\bibitem[{{Priest} \& {D{\'e}moulin}(1995)}]{Priest1995JGR...10023443P}
{Priest}, E.~R., \& {D{\'e}moulin}, P. 1995, \jgr, 100, 23443,
  \dodoi{10.1029/95JA02740}

\bibitem[{Qiu {et~al.}(2004)Qiu, Wang, Cheng, \&
  Gary}]{Qiu2004ApJ_ReconnectionRate}
Qiu, J., Wang, H., Cheng, C.~Z., \& Gary, D.~E. 2004, The Astrophysical
  Journal, 604, 900, \dodoi{10.1086/382122}

\bibitem[{Resnick(2007)}]{Resnick2007HeavyTail}
Resnick, S.~I. 2007, Heavy-Tail Phenomena: Probabilistic and Statistical
  Modeling (New York: Springer), \dodoi{10.1007/978-0-387-75953-1}

\bibitem[{{Rimmele} {et~al.}(2020){Rimmele}, {Warner}, {Keil}, {Goode},
  {Kn{\"o}lker}, {Kuhn}, {Rosner}, {McMullin}, {Casini}, {Lin}, {W{\"o}ger},
  {von der L{\"u}he}, {Tritschler}, {Davey}, {de Wijn}, {Elmore}, {Fehlmann},
  {Harrington}, {Jaeggli}, {Rast}, {Schad}, {Schmidt}, {Mathioudakis},
  {Mickey}, {Anan}, {Beck}, {Marshall}, {Jeffers}, {Oschmann}, {Beard},
  {Berst}, {Cowan}, {Craig}, {Cross}, {Cummings}, {Donnelly}, {de Vanssay},
  {Eigenbrot}, {Ferayorni}, {Foster}, {Galapon}, {Gedrites}, {Gonzales},
  {Goodrich}, {Gregory}, {Guzman}, {Guzzo}, {Hegwer}, {Hubbard}, {Hubbard},
  {Johansson}, {Johnson}, {Liang}, {Liang}, {McQuillen}, {Mayer}, {Newman},
  {Onodera}, {Phelps}, {Puentes}, {Richards}, {Rimmele}, {Sekulic}, {Shimko},
  {Simison}, {Smith}, {Starman}, {Sueoka}, {Summers}, {Szabo}, {Szabo},
  {Wampler}, {Williams}, \& {White}}]{rimmele20a}
{Rimmele}, T.~R., {Warner}, M., {Keil}, S.~L., {et~al.} 2020, \solphys, 295,
  172, \dodoi{10.1007/s11207-020-01736-7}

\bibitem[{Savage {et~al.}(2012)Savage, McKenzie, \&
  Reeves}]{Savage2012_ApJL_SADsReinterpretation}
Savage, S.~L., McKenzie, D.~E., \& Reeves, K.~K. 2012, The Astrophysical
  Journal Letters, 747, L40, \dodoi{10.1088/2041-8205/747/2/L40}

\bibitem[{{Shibata} \& {Tanuma}(2001)}]{Shibata_etal2001}
{Shibata}, K., \& {Tanuma}, S. 2001, Earth, Planets, and Space, 53, 473,
  \dodoi{10.1186/BF03353258}

\bibitem[{Song {et~al.}(2012)Song, Kong, Chen, Li, Li, Feng, \&
  Xia}]{Song2012_SolPhys_CMEBlobs}
Song, H.-Q., Kong, X.-L., Chen, Y., {et~al.} 2012, Solar Physics, 276,
  \dodoi{10.1007/s11207-011-9848-9}

\bibitem[{{Sturrock}(1966)}]{Sturrock_1966}
{Sturrock}, P.~A. 1966, \nat, 211, 695, \dodoi{10.1038/211695a0}

\bibitem[{Suzuki \& Abe(1985)}]{SuzukiAbe1985}
Suzuki, S., \& Abe, K. 1985, Computer Vision, Graphics, and Image Processing,
  30, 32, \dodoi{10.1016/0734-189X(85)90016-7}

\bibitem[{Takasao {et~al.}(2016)Takasao, Asai, Isobe, \&
  Shibata}]{Takasao2016PlasmoidMotions}
Takasao, S., Asai, A., Isobe, H., \& Shibata, K. 2016, The Astrophysical
  Journal, 828, 103, \dodoi{10.3847/0004-637X/828/2/103}

\bibitem[{{Tamburri} {et~al.}(2025){Tamburri}, {Kazachenko}, {Cauzzi},
  {Kowalski}, {French}, {Yadav}, {Evans}, {Notsu}, {Corchado-Albelo},
  {Reardon}, \& {Tritschler}}]{Tamburri2025}
{Tamburri}, C.~A., {Kazachenko}, M.~D., {Cauzzi}, G., {et~al.} 2025, \apjl,
  990, L3, \dodoi{10.3847/2041-8213/adf95e}

\bibitem[{{Thoen Faber} {et~al.}(2025){Thoen Faber}, {Joshi}, {Rouppe van der
  Voort}, {Wedemeyer}, {Fletcher}, {Aulanier}, \&
  {N{\'o}brega-Siverio}}]{Faber2025}
{Thoen Faber}, J., {Joshi}, R., {Rouppe van der Voort}, L., {et~al.} 2025,
  \aap, 693, A8, \dodoi{10.1051/0004-6361/202452370}

\bibitem[{Uzdensky {et~al.}(2010)Uzdensky, Loureiro, \&
  Schekochihin}]{Uzdensky2010PRL}
Uzdensky, D.~A., Loureiro, N.~F., \& Schekochihin, A.~A. 2010, Phys. Rev.
  Lett., 105, 235002, \dodoi{10.1103/PhysRevLett.105.235002}

\bibitem[{{Vassilicos} \& {Hunt}(1991)}]{Vassilicos1991}
{Vassilicos}, J.~C., \& {Hunt}, J.~C.~R. 1991, Proceedings of the Royal Society
  of London Series A, 435, 505, \dodoi{10.1098/rspa.1991.0158}

\bibitem[{Warren {et~al.}(2018)Warren, Brooks, Ugarte-Urra, Reep, Crump, \&
  Doschek}]{Warren2018_ApJ_CurrentSheetSpectroscopy}
Warren, H.~P., Brooks, D.~H., Ugarte-Urra, I., {et~al.} 2018, The Astrophysical
  Journal, 854, 122, \dodoi{10.3847/1538-4357/aaa9b8}

\bibitem[{Welzl(1991)}]{Welzl1991}
Welzl, E. 1991, in LNCS, Vol. 555, New Results and New Trends in Computer
  Science (Springer), 359--370, \dodoi{10.1007/BFb0038202}

\bibitem[{{Wyper} {et~al.}(2021){Wyper}, {Antiochos}, {DeVore}, {Lynch},
  {Karpen}, \& {Kumar}}]{Wyper2021}
{Wyper}, P.~F., {Antiochos}, S.~K., {DeVore}, C.~R., {et~al.} 2021, \apj, 909,
  54, \dodoi{10.3847/1538-4357/abd9ca}

\bibitem[{{Wyper} \& {DeVore}(2016)}]{wyper2016ApJ...820...77W}
{Wyper}, P.~F., \& {DeVore}, C.~R. 2016, \apj, 820, 77,
  \dodoi{10.3847/0004-637X/820/1/77}

\bibitem[{{Wyper} {et~al.}(2016){Wyper}, {DeVore}, {Karpen}, \&
  {Lynch}}]{wyper2016ApJ...827....4W}
{Wyper}, P.~F., {DeVore}, C.~R., {Karpen}, J.~T., \& {Lynch}, B.~J. 2016, \apj,
  827, 4, \dodoi{10.3847/0004-637X/827/1/4}

\bibitem[{{Wyper} {et~al.}(2024){Wyper}, {Lynch}, {DeVore}, {Kumar},
  {Antiochos}, \& {Daldorff}}]{Wyper2024}
{Wyper}, P.~F., {Lynch}, B.~J., {DeVore}, C.~R., {et~al.} 2024, \apj, 975, 168,
  \dodoi{10.3847/1538-4357/ad7941}

\bibitem[{{Wyper} \& {Pontin}(2014)}]{Wyper2014a}
{Wyper}, P.~F., \& {Pontin}, D.~I. 2014, \phpl, 21, 082114,
  \dodoi{10.1063/1.4893149}

\bibitem[{{Wyper} \& {Pontin}(2021)}]{Wyper_Pontin2021ApJ...920..102W}
---. 2021, \apj, 920, 102, \dodoi{10.3847/1538-4357/ac1943}

\bibitem[{{Yadav} {et~al.}(2025){Yadav}, {Kazachenko}, {Cauzzi}, {Tamburri},
  {Corchado}, \& {French}}]{Yadav2025}
{Yadav}, R., {Kazachenko}, M.~D., {Cauzzi}, G., {et~al.} 2025, \apj, 989, 183,
  \dodoi{10.3847/1538-4357/adf4c1}

\bibitem[{{Zhang}(2024)}]{Zhang2024}
{Zhang}, Q. 2024, Reviews of Modern Plasma Physics, 8, 7,
  \dodoi{10.1007/s41614-024-00144-9}

\end{thebibliography}

\end{document}